\documentclass[floatfix,twocolumn,aps,prd,showpacs,amsmath,amssymb]{revtex4}
\usepackage{graphicx}
\usepackage{hyperref}
\hypersetup{colorlinks=false}
\hypersetup{linktocpage}
\newcommand{\A}{D}  
\newcommand{\B}{C}  
\newcommand{\beq}{\begin{equation}}
\newcommand{\eeq}{\end{equation}}

\newcommand{\uup}{u^{\text{up}}}
\newcommand{\uin}{u^{\text{in}}}
\newcommand{\Ain}{A^{\text{in}}_{m \omega}}
\newcommand{\Aout}{A^{\text{out}}_{m \omega}}
\newcommand{\Bin}{B^{\text{in}}_{m \omega}}
\newcommand{\Bout}{B^{\text{out}}_{m \omega}}
\newcommand{\omtil}{\tilde{\omega}}
\newcommand{\nn}{\nonumber}
\newcommand{\Lc}{l_c}
\newcommand{\lc}{l_c}
\newcommand{\rc}{r_c}
\newcommand{\sinc}{\mathop{\text{sinc}}}
\usepackage{color}

\newcommand{\sigabs}{\sigma_{\text{abs}}}
\newcommand{\Vgeo}{V_{\text{geo}}}

\newcommand{\uint}{\mathcal{X}}

\begin{document}
\title{Resonances of a rotating black hole analogue}

\author{Sam R. Dolan}
\email{s.dolan@soton.ac.uk}
\affiliation{School of Mathematics, University of Southampton,
Highfield, Southampton SO17 1BJ, United Kingdom}

\author{Leandro A. Oliveira}
\email{laoliveira@ufpa.br}
\affiliation{Faculdade de
F\'\i sica, Universidade Federal do Par\'a, 66075-110, Bel\'em,
Par\'a, Brazil}

\author{Lu\'\i s C. B. Crispino}
\email{crispino@ufpa.br}
\affiliation{Faculdade de F\'\i sica, Universidade Federal do
Par\'a, 66075-110, Bel\'em, Par\'a, Brazil}

\date{\today}
\begin{abstract}
Under certain conditions, sound waves in a fluid may be governed by a Klein-Gordon equation on an `effective spacetime' determined by the background flow properties. Here we consider the draining bathtub: a circulating, draining flow whose effective spacetime shares key features with the rotating black hole (Kerr) spacetime. We present a complete investigation of the role of quasinormal (QN) mode and Regge pole (RP) resonances of this system. First, we simulate a perturbation in the time domain by applying a finite-difference method, to demonstrate the ubiquity of `QN ringing'. Next, we solve the wave equation in the frequency domain with the continued-fraction method, to compute QN and RP spectra numerically. We then explore the geometric link between (prograde and retrograde) null geodesic orbits on the spacetime, and the properties of the QN/RP spectra. We develop a `geodesic expansion' method which leads to asymptotic expressions (in inverse powers of mode number $m$) for the spectra, the radial functions and the residues. Next, the role of the Regge poles in scattering and absorption processes is revealed through application of the complex angular momentum method. We elucidate the link between the Regge poles and oscillations in the absorption cross section. Finally, we show that Regge poles provide a neat explanation for `orbiting' oscillations seen in the scattering cross section.
\end{abstract}


\maketitle

\section{Introduction}
Black holes -- `trapped' regions of spacetime -- are a key element of Einstein's theory of general relativity. Although perhaps once viewed as mathematical curiosities, astronomers have now compiled a range of compelling evidence for their existence. Black holes are a key ingredient in modern theories of galaxy formation, quasars, accretion disks, gamma ray bursts and supernovae. Yet, even if black holes were nothing more than a theorist's `thought experiment', they would still have provoked the development of theoretical physics. In the 1970s, Hawking and others showed that quantum field theory in curved spacetime implies that black holes are not completely black: they must radiate thermally, with a negative heat capacity. Furthermore, black holes seem to have a well-defined entropy, that scales with the horizon area \cite{haw1}. This realisation has inspired myriad (and ongoing) attempts to consistently combine relativity and field theory in the strong field regime. 

There is more to black holes than Hawking radiation. A key property of a black hole is that it bends and traps light. Light rays may orbit a black hole in the vicinity of a `photon sphere' which lies somewhat outside the horizon (at $r=3 r_h / 2$ for the Schwarzschild BH, where $r_h$ is the horizon radius). The existence of an unstable photon orbit gives rise to various interesting effects, such as the strong-field lensing of light from a distant source passing close to a black hole. In particular, the photon orbit is intimately linked to characteristic `damped resonances' that appear when waves interact with a black hole. Mathematically, damped resonances are manifest as poles in the scattering matrix $S$. The poles occur at complex frequencies and at complex angular momenta, and the corresponding modes are known as quasinormal (QN) and Regge pole (RP) modes, respectively \cite{Kokkotas-Schmidt,Nollert,Ferrari-Gualtieri,Berti-Cardoso-Starinets,Konoplya-Zhidenko:2011}. In this paper, we investigate QN and RP modes and their physical consequences, in the setting of a simple rotating `black hole analogue' \cite{Visser}. 

It seems unlikely that we will ever study black holes directly in the laboratory. Yet, as Unruh  \cite{Unruh} noted three decades ago, we may study analogues: artificial systems (in various media) which exhibit some key kinematic features of black holes \cite{Novello-Visser-Volovik}. For example, sound waves in fluid flows which are inviscid, irrotational, and barotropic are governed by the same wave equation as a scalar field in a curved space-time, namely \cite{Visser}
\beq
 \Box \Phi = \frac{1}{\sqrt{|g|}} \partial_\mu \left( \sqrt{|g|} g^{\mu \nu} \partial_\nu \Phi \right) = 0 , \label{kg}
\eeq
where here $g_{\mu \nu}$ is the \emph{effective metric} (with inverse $g^{\mu \nu}$ and determinant $g$). Note that in this context, $g_{\mu \nu}$ depends algebraically on the local properties of the fluid flow, and it does not (in general) represent a solution of the Einstein equations. Nevertheless, it is an intriguing prospect that, by studying sound waves on a background flow, one may understand better the propagation of fields on a curved spacetime. In recent years, a wide range of black hole analogues in various media have been proposed and, indeed, studied, in the laboratory \cite{Garay-2002, Garay 2, Schutzhold-Unruh-2004, Belgiorno, Giovanazzi}. A surge of recent experimental activity appears to be bearing fruit, as evidenced by a recent claim of experimental observation of correlations related to Hawking radiation in a wave-tank \cite{Wein}. In this experiment, instead of a black-hole analogue, a white-hole analogue is used. A further example of a simple `white-hole' analogue in fluids is the so-called circular hydraulic jump÷\cite{Jannes1}.

One of the simplest analogue models is the so-called {\it draining bathtub} (or draining vortex), described in \cite{Visser}: a two-dimensional circulating flow with a sink at the origin. In 2002, Sch\"utzhold and Unruh \cite{Schutzhold-Unruh-2002} described a possible experimental realization of the `bathtub' idea, wherein gravity waves propagate in a flowing fluid in a shallow basin of varying height $h(r)$. Non-dispersive long-wavelength perturbations are governed by an effective geometry with line element $ds^2 = g_{\mu \nu} dx^\mu dx^\nu$, where
\beq
ds^2 = -c^2 d\tilde{t}^2 + \left( dr + \frac{D d\tilde{t}}{r} \right)^2 + \left(r d \tilde{\phi} - \frac{C d\tilde{t}}{r} \right)^2. \label{le1}
\eeq
Here the constants of circulation ($C$) and draining ($D$) \cite{footnote} 
relate to the background flow velocity $\mathbf{v}_0$ of the fluid, namely
\beq
\mathbf{v}_0 = - D \hat{r} / r + C \hat{\phi} / r. 
\eeq
We assume that $D > 0$, so that the system acts as a black, rather than white, hole. In the model described in Ref.~\cite{Schutzhold-Unruh-2002}, the speed of the perturbation in the fluid $c$ is set by $c^2 = {a_g} h_\infty$, where $a_g$ is the acceleration due to gravity and $h_\infty$ is the height of the fluid far from the centre. Note that we assume we are within the linear dispersion regime, so that the perturbations propagate with a constant speed (i.e.~$c$ is independent of frequency). The analogue event horizon (where the inward flow rate exceeds $c$) lies at $r_h = \A / c$, and the analogue ergosphere (where the flow becomes supersonic $|\mathbf{v}_0| \geq c$) has a boundary at $r_e = \sqrt{\B^2 + \A^2} / c$ \cite{Visser}. Small perturbations $\delta \mathbf{v}$ to the flow, $\mathbf{v} = \mathbf{v}_0 + \delta{\mathbf{v}}$, may be expressed in terms of a gradient of a potential $\delta \mathbf{v} = - \nabla \Phi$, and the potential field $\Phi$ satisfies the Klein-Gordon equation (\ref{kg}) with effective metric (\ref{le1}). Henceforth we set the speed of the perturbation equal to unity $(c = 1)$.  

There are several key motivations for considering the draining bathtub (DBT). Firstly, as described above, the DBT may perhaps be realized in the laboratory. Secondly, the DBT provides a useful `toy model' for the most astrophysically-relevant black hole, i.e.~the Kerr solution. For example, both the DBT and the Kerr solution possess a horizon and an ergosphere, and can exhibit superradiance; but whereas the angular momentum of the (non-naked) Kerr black hole is constrained, $J \le M^2$, the angular momentum of the DBT is (in principle) unbounded. This follows as a consequence of their differing symmetry: the DBT is cylindrically symmetric, whereas the Kerr solution is axially symmetric. A third reason is simplicity: the DBT is arguably the simplest asymptotically flat \emph{rotating} spacetime that can be envisaged (see also the cosmic string \cite{Xanthopoulos}). It serves as a testing ground for developing calculation methods that can be extended to Kerr geometry.

Given such motivations, it is no surprise to find that the spectrum of quasinormal modes of the DBT has already received some attention \cite{Berti-Cardoso-Lemos, Cardoso-Lemos-Yoshida}. On the other hand, the Regge pole spectrum has not been considered. Given the close relationship between QN and RP modes, we believe that a comprehensive study of the resonances of the DBT is now justified. Here we can go many steps beyond existing work to show: (i) how quasinormal resonances arise in time-domain simulations of a small perturbation in the flow, (ii) how QN and RP resonances are closely related to the properties of the co- and counter-rotating null orbits, (iii) how complex angular momentum methods \cite{Andersson-Thylwe-1994, Andersson-1994, Decanini-Folacci-Jensen} may be applied to compute absorption and scattering cross sections \cite{ODC, DOC2}, (iv) how the fine-structure of the absorption cross section is related to the Regge pole spectrum \cite{DEFF, DFR}, and (v) how `orbiting' oscillations in the scattering cross section are also related to Regge pole spectrum. 

The remainder of this paper is structured as follows. In Sec.~\ref{DBT} we study perturbations and null geodesics in the DBT effective spacetime. In Sec.~\ref{resonances} we evolve Gaussian initial data in the time domain and identify the quasinormal mode ringing signal. We apply the continued-fraction method to obtain the frequency spectrum of the DBT QN modes. We extend the geodesic expansion method of Ref.~\cite{Dolan-Ottewill, Dolan} and apply it to find the QN modes of the DBT. We validate the expansion method formulae by comparing with numerical results from the continued-fraction method. In Sec.~\ref{RPs} we extend the expansion method to find approximations for the Regge poles, and again check against numerical data. In Sec.~\ref{sec:CAM} we harness the power of the Complex Angular Momentum method to  understand the key features of the DBT absorption and scattering cross sections.  In Sec.~\ref{subsec:abs-csec} we show that the absorption cross section $\sigabs(\omega)$ of the DBT \cite{ODC} may be expressed in terms of the Regge poles \cite{DEFF, DFR}, and we derive a simple `geometric' approximation for the cross section. In Sec.~\ref{subsec:scat-csec} we show that `orbiting' oscillations which arise in the scattering cross section are also related to the Regge poles. We extend the geometric expansion method (in Appendix A) to obtain an approximation for the \emph{residues} of the Regge poles, which we use to derive a simple geometric approximation for `orbiting' at large scattering angles. We validate our approximations against numerical data. We conclude with our final remarks in Sec.~\ref{conclusions}.

\section{The Draining Bathtub\label{DBT}}
The draining bathtub was briefly described in the previous section, and we address the reader to Ref. \cite{Visser} for more details. In the lab-based  coordinates $\{\tilde{t}, r, \tilde{\phi} \}$, the effective geometry is described by the line element (\ref{le1}). Following \cite{Berti-Cardoso-Lemos}, it is convenient to introduce an alternative coordinate system $\{t, r, \phi \}$
via $d t = d \tilde{t} - \A dr / (r f) $, $d \phi = d \tilde{\phi} - \B \A dr / (r^3 f)$, 
 with $\phi(r \to \infty) \to \tilde{\phi}$ and
\beq
f(r) = 1 - \A^2/r^2 .
\eeq
The line element~(\ref{le1}) then takes the form
\beq
ds^2 = -f(r) d t^2 + f(r)^{-1} dr^2 + \left(r d\phi - \B dt / r \right)^2 .
\label{le2}
\eeq
Henceforth we will work with these (non-lab based) coordinates exclusively. 

\subsection{Perturbations}
Small perturbations to the background flow, $\delta \mathbf{v} = - \nabla \Phi$, are governed by the Klein-Gordon equation (\ref{kg}) with line element (\ref{le2}). Let us now decompose $\Phi$ in azimuthal modes, namely
\beq
\Phi(t,r,\phi) = \frac{1}{\sqrt{r}} \sum_{m=-\infty}^{\infty} \psi_m(t, r) e^{i m \phi} . \label{modesum}
\eeq
Since $\Phi$ is a real field, the symmetry relation $\psi_m^* = \psi_{-m}$ follows. Inserting Eq.~(\ref{modesum}) into Eq.~(\ref{kg}) leads to the wave equation
\beq
\left[ -\left(\frac{\partial}{\partial t} + \frac{i C m}{r^2} \right)^2 + \frac{\partial^2}{\partial r_*^2} - V_m(r) \right] \psi_m(t,r) = 0 ,  \label{waveeq}
\eeq
where
\beq
V_m(r) = f(r) \left[ \frac{m^2 - 1/4}{r^2} + \frac{5\A^2}{4r^4} \right] ,
\eeq
and the tortoise coordinate is defined by $dr_\ast = f^{-1} dr$, or, explicitly,
\beq
r_* = r + \frac{\A}{2} \ln \left| \frac{r - \A}{r + \A} \right| .
\eeq


A perturbation of compact support in the vicinity of the hole satisfies the boundary conditions
\begin{eqnarray}
\lim_{r_* \rightarrow -\infty} \left[ \partial_t + i C m / D^2 - \partial_{r_*}  \right] \psi_m &=& 0, \\
\lim_{r_* \rightarrow +\infty} \left[ \partial_t + \partial_{r_*} \right] \psi_m &=& 0. 
\end{eqnarray}

 \subsection{Geodesics\label{subsec:geodesics}}
According to the eikonal approximation, very short-wavelength perturbations propagate along null geodesics of the effective spacetime (\ref{le2}), i.e.
\beq
\Phi \sim \exp \left( -i k_{\mu} x^{\mu} \right), \quad k_{\mu} k^{\mu} = 0 , \quad k^{\mu} k_{\nu ; \mu} = 0 .
\eeq
Hence, by investigating the null geodesics of the line element~(\ref{le2}) we may understand high-frequency wave propagation. Let us consider a geodesic with tangent vector $k^{\mu} = ( \dot{t}, \dot{r}, \dot{\phi}) $, where the overdot denotes differentiation with respect to an affine parameter. Geodesics have two constants of motion, i.e.,~energy and angular momentum
\begin{eqnarray}
E &=& \left( 1 - \frac{\A^2+\B^2}{r^2} \right) \dot{t} + \B\dot{\phi},
\nonumber
\label{E} \\
L &=& -\B \dot{t} + r^2 \dot{\phi},
\nonumber
\label{L}
\end{eqnarray}
respectively. For a null geodesic ($k^\mu k_\mu = 0$) we may write the `energy equation' as
\begin{equation}
\dot{r}^2 + \Vgeo(r) = E^2,
\label{geo_eq}
\end{equation}
where
\begin{equation}
\Vgeo(r) = \left(1 - \frac{\A^2+\B^2}{r^2} \right) \frac{L^2}{r^2}
+\frac{2 \B L E}{r^2}.
\label{V}
\end{equation}

Let us consider a null geodesic impinging from infinity. Its `impact parameter' $b$ may be defined as the perpendicular distance (measured at infinity) between the geodesic and a parallel line that passes through the origin,
$$b\equiv L/E + \B .$$
Here $b$ is defined as a displacement that may take either sign, allowing us to distinguish between co-rotating ($+$) and counter-rotating ($-$) geodesics. Now, if $|b|$ is large, the geodesic will be scattered; if $|b|$ is small, the geodesic will be absorbed (i.e.~it will pass through the horizon). Between these regimes is a `critical' geodesic, which is neither scattered nor absorbed but instead ends in perpetual orbit at $r=r_c^\pm$.   By solving simultaneous conditions $\Vgeo(r_c) = E^2$ and $d \Vgeo / dr(r_c) = 0$ we find a pair of `critical' geodesics given by
\begin{eqnarray}
b_{c}^{\pm} &=& -\B \pm 2\sqrt{\A^2 + \B^2},   \label{bc} \\
r_{c}^{\pm}  &=& \left( \sqrt{\A^2 + \B^2} \, \left| b_c^\pm - \B \right|  \right)^{1/2}  \label{rc}.
\end{eqnarray}
Note that $b_c^{-}$ is defined to be negative. For $C > 0$, geodesics with $b>\B$ ($L>0$) co-rotate 
with the system, whereas geodesics with $b<\B$ ($L<0$) counter-rotate.


It is natural to define $l \equiv b - \B$, which is positive 
for co-rotating geodesics and negative for the counter-rotating ones,
so that:
\begin{equation}
\Lc^{\pm} \equiv b^{\pm}_{c}-\B = -2\B \pm 2\sqrt{\A^2 + \B^2}.  
\label{lc}
\end{equation}
Trajectories of null geodesics impinging from spatial infinity upon a draining bathtub are illustrated in Ref. \cite{ODC}. 




\section{Quasinormal Mode Resonances\label{resonances}}
With these preliminaries established, let us now turn our attention to quasinormal (QN) modes. We start in Sec.~\ref{subsec:qn-timedomain} by demonstrating QN resonances in a perturbation encroaching upon a black hole, by evolving Gaussian initial data in the time domain. In Sec.~\ref{sec:qn-freq} we recap the theory of QN modes as poles of the Green function in the frequency domain, and in Sec.~\ref{sec:qn-spectrum} we describe the symmetries of the spectrum and provide an  exact solution for the non-resonant $m=0$ mode. Next, we apply two frequency-domain methods to determine the spectrum: the numerical method of \cite{Cardoso-Lemos-Yoshida}, and the `geodesic expansion' method of \cite{Dolan-Ottewill,Dolan}. We validate our results, compare with the Kerr spectrum, and give a geometric interpretation. Henceforth we set $r_h = \A  = 1$ for convenience, unless otherwise stated.

  \subsection{Quasinormal modes in the time domain\label{subsec:qn-timedomain}}

To study the phenomenon of QN mode ringing, let us consider the evolution of a small perturbation to the system. For concreteness, we take an initial condition of the form
\begin{eqnarray}
\psi_m(t = 0, r_\ast) &=& \exp\left[ - (r_\ast - r_{\ast 0})^2 / (2 \sigma^2) \right], \\
\partial_t \psi_m(t=0, r) &=& 0, \label{eq:ic}
\end{eqnarray}
where $r_{\ast 0}$ and $\sigma^2$ are arbitrary constants. Now, we simulate the evolution of this initial perturbation by applying finite-difference (FD) methods to the wave equation (\ref{waveeq}). There are many possible FD methods; we chose the `Method of Lines' (described e.g.~in Ref.~\cite{Schiesser} and Sec.~4.2 of Ref.~\cite{Rinne}), using second-order differencing on spatial slices and the fourth-order Runge-Kutta method to advance in time. This method was chosen primarily for its good stability properties: it is numerically stable provided that the time step $\tau$ is small enough (typically we took $\tau = h / 2$ for our evolutions, where $h$ is the spatial grid spacing). We made the spatial domain large enough in $r_\ast$ so that the `physical' perturbation did not encounter the spatial boundaries during the simulation run.

\begin{figure}
   \includegraphics[width=8.2cm]{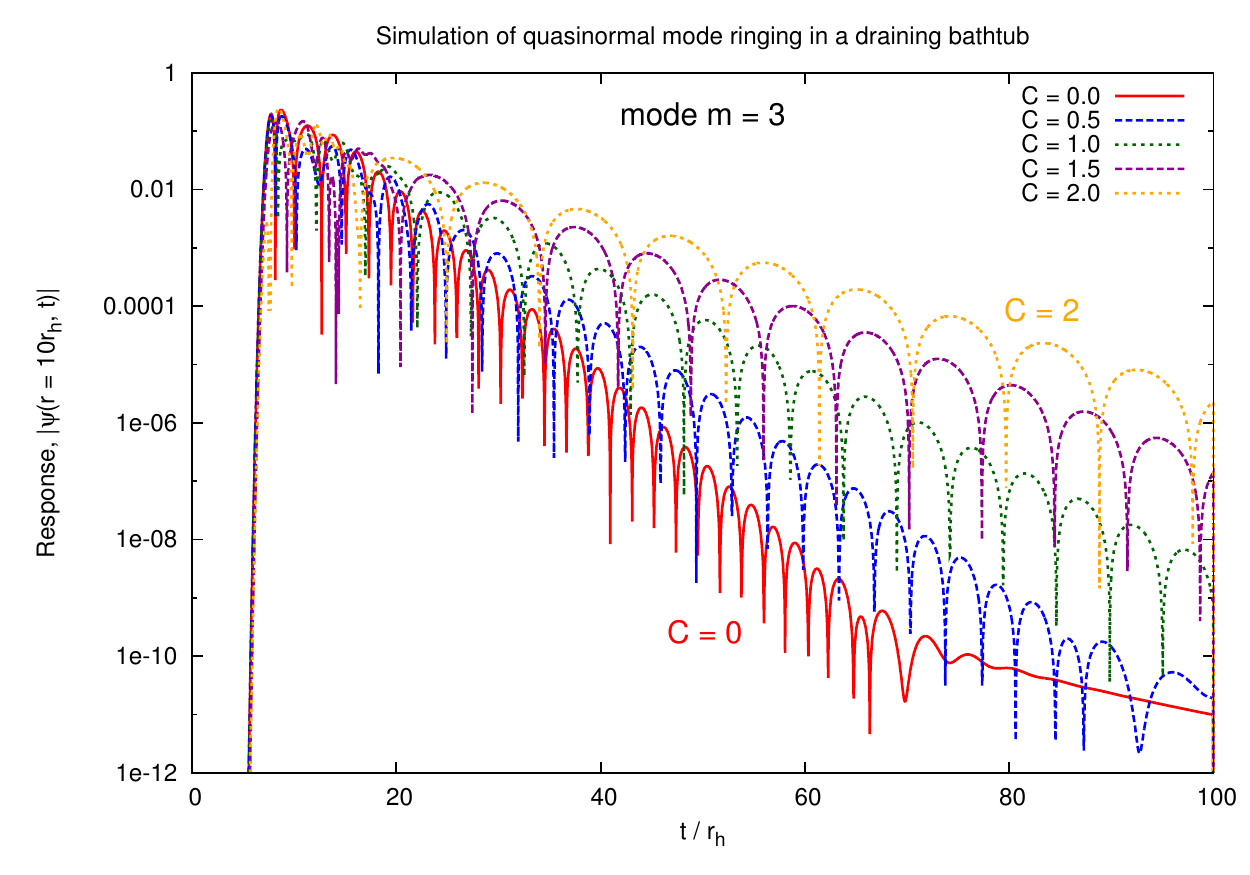}
  \caption{Illustrating the `quasinormal ringing' response as a function of time $t$ of a draining bathtub to a Gaussian initial perturbation. Here the value of $\psi(t, r)$ at a fixed radius $r=10r_h$ is shown, for various values of the `circulation' $C$ of the bathtub, $C=0 \ldots 2$. As the circulation rate increases, the ringing frequency and the damping rate decrease. }
   \label{fig:timedomain-ringing}
  \end{figure}

Typical results are given in Fig.~\ref{fig:timedomain-ringing}, which shows the reponse $\psi_m(t,r)$ at fixed radius $r = 10r_h$ as a function of time, for an initial condition of the form (\ref{eq:ic}). The logarithmic scale on the vertical axis shows that, at intermediate times, the response is apparently dominated by exponentially-damped ringing. This response is typical of systems with an unstable geodesic orbit (or equivalently a peak in the potential barrier). Ringing may be understood in terms of QN modes with characteristic complex frequencies $\omega_{mn}^{\pm}$ which depends on parameters of the system ($C$ and $D$) rather than details of the initial perturbation.

\begin{figure}
    \includegraphics[width=8.2cm]{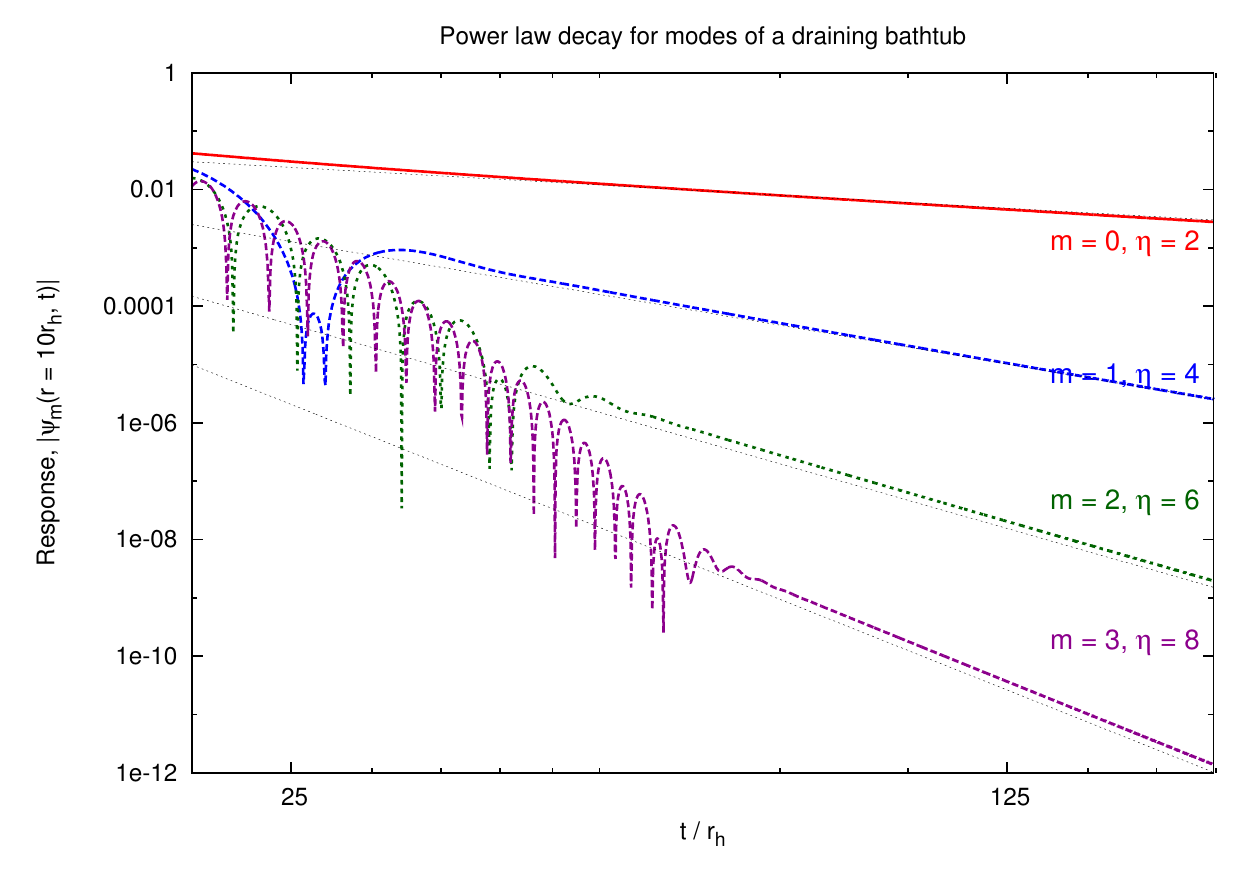}
  \caption{Log-log plot illustrating `power law decay' in modes $m=0 \ldots 3$ for a perturbation of a draining bathtub. At late times, the perturbation decays as $\psi \propto (t/r_h)^{-\eta}$, where $\eta = 2 |m| + 1$ for generic (in this case, Gaussian) initial data, and $\eta = 2|m| + 2$ for time-symmetric initial data, $\partial_t \psi(t=0, r) = 0$, which is shown here. The decay rate is independent of the circulation $C$.}
   \label{fig:timedomain-powerlaw}
  \end{figure}

The time-domain signal in Fig.~\ref{fig:timedomain-ringing} clearly shows the imprint of the least-damped QN mode. In later sections, we find that (for $m \neq 0$) the least-damped mode is the retrograde (rather than prograde) mode. Fig.~\ref{fig:timedomain-ringing} shows that the damping rate and ringing frequency of this mode decreases as the circulation rate $C$ increases; in Sec.~\ref{subsec:geometric} we relate this behavior to the properties of geodesics. The ringing frequency increases with mode number $m$. Figure~\ref{fig:timedomain-powerlaw} shows that, at very late times, the signal is dominated by a power law decay, 
$$
\psi(t, r) \sim t^{-\eta},
$$
where
\beq
\eta = \begin{cases} 2 |m| + 1, \quad & \partial_t \psi |_{t=0} \neq 0, \\ 2 |m| + 2, & \partial_t \psi |_{t=0} = 0 . \end{cases}
\label{power-law-index}
\eeq
In other words, the decay is one power of $t$ more rapid in the special case of time-symmetric initial data [such as Eq.~(\ref{eq:ic})]. For further discussion of power-law decay in the DBT context, see Sec.~IIE in Ref.~\cite{Berti-Cardoso-Lemos}.

  \subsection{Quasinormal modes in the frequency domain\label{sec:qn-freq}}
It is well-known that the origin of QN ringing and power law decay may be understood by considering the decomposition of the Green's function in the frequency domain (see for example \cite{Leaver-1986, Andersson-1997}). Let us briefly recap the argument here. A formal solution of Eq.~(\ref{waveeq}) with arbitrary initial condition $\psi_m^0(x) \equiv \psi_m(t=0, x)$, $\dot{\psi}_m^0(x) \equiv  \partial_t \psi_m(t,x)|_{t=0}$ may be written as
\begin{eqnarray}
\psi_m(t,x) =\int \left[ G(x, x', t) \dot{\psi}_m^0(x') + \right. \nonumber \\
\left. \partial_t G\left(x, x', t \right) \psi_m^0 \left(x' \right) \right] dx',
\end{eqnarray}
where $x=r_\ast$, $x' = r_\ast'$ and $G$ is the (retarded) Green's function defined by
\beq
\left[ \frac{\partial^2}{\partial r_*^2} - \left(\frac{\partial}{\partial t} + \frac{i C m}{r^2} \right)^2 - V_m(r) \right] G = \delta(t) \delta(x - x'),
\eeq
and $G = 0$ for $t < 0$. The Green's function may be written in terms of an inverse Fourier transform
\beq
G(x,x', t) = \frac{1}{2\pi} \int_{-\infty + i \zeta}^{\infty + i \zeta} \hat{G}(x,x', \omega) e^{- i \omega t} d\omega, \label{GinvFT}
\eeq
with $\zeta$ positive and real. The frequency-domain function $\hat{G}$ is constructed in the following way:
\beq
\hat{G}(x,x', \omega) = -\frac{1}{2 i \omega \Ain} \begin{cases} \uin_{m\omega} (x) \uup_{m\omega}(x'), \quad & x \le x', \\ \uin_{m \omega} (x') \uup_{m \omega} (x), \quad & x \ge x'. \end{cases} \label{GFfreq}
\eeq
Here $\uin_{m\omega}(r)$ (`in') and $\uup_{m\omega}$ (`up') are solutions of the homogeneous equation
\beq
\left[ \frac{d^2}{d r_\ast^2} + \left(\omega - \frac{C m}{r^2} \right)^2 - V_m(r) \right] u(r) = 0 \label{radeq}
\eeq
that are, respectively, ingoing at the horizon and outgoing at infinity, satisfying the following asymptotic boundary conditions
\beq
\uin_{m\omega}(r_\ast) \sim \begin{cases} e^{-i\omtil r_\ast},  & r_\ast \rightarrow -\infty , \\  \Aout e^{+i \omega r_\ast} + \Ain e^{-i \omega r_\ast} , & r_\ast \rightarrow +\infty,
\end{cases}
\label{uin}
\eeq
and
\beq
\uup_{m\omega}(r_\ast) \sim  \begin{cases} \Bout e^{+i \omtil r_\ast} + \Bin e^{-i \omtil r_\ast},  & r_\ast \rightarrow -\infty , \\  e^{+i \omega r_\ast},  & r_\ast \rightarrow +\infty, 
\end{cases}
\eeq
where $\tilde{\omega} \equiv \omega - mC / D^2$. By considering the Wronskian, one may establish certain relationships between these constants, for example, $\omega \Ain = \omtil \Bout$. 

The frequency-domain Green's function $\hat{G}$ given in Eq.~(\ref{GFfreq}) has poles at frequencies for which $\Ain = 0$. Such poles do not lie on the real frequency axis, but rather in the lower half of the complex frequency plane. For $t > 0$, the contour of integration in Eq.~(\ref{GinvFT}) may be closed in the lower half-plane, as described e.g.~in Refs.~\cite{Leaver-1986, Andersson-1997}, enclosing the poles. By Cauchy's theorem there arises a sum over residues of these poles, which is known as a `QN mode sum'. The QN mode sum manifests itself as the damped ringing response seen e.g. in~Fig.~\ref{fig:timedomain-ringing}. Furthermore, there exists a branch point in $\hat{G}$ at zero frequency, which necessitates a branch cut along the negative imaginary axis. The integral of $\hat{G}$ along either side of the branch cut is associated with the power-law decay at late times, as observed in Fig.~\ref{fig:timedomain-powerlaw}. With this in mind, let us now consider the QN spectrum, i.e.~the set of frequencies defined by $\Ain = 0$.

\subsection{The quasinormal mode spectrum\label{sec:qn-spectrum}}
\subsubsection{The $m = 0$ mode}
The $m=0$ mode is isotropic and independent of $C$, and furthermore, Eq.~(\ref{radeq}) has a simple closed-form solution in this case. The `in' mode is 
\beq
\uin_{0\omega}(r) = r^{1/2} e^{-i\beta} (i \beta)^{i \beta} \Gamma(1 - i\beta) I_{-i\beta}(i \omega r f^{1/2}) , \label{uin-m0}
\eeq
where $\beta \equiv \omega \A$ and $I_{\nu}(z)$ is a modified Bessel function of the first kind \cite{Grad}. In this case, the boundary condition constants are
\begin{eqnarray}
A^{\text{out}}_{0\omega} &=& \left( 2 \pi i \beta \right)^{-1/2} e^{-i \beta} (i \beta)^{i \beta} \Gamma(1 - i\beta),  \label{Aout-m0} \\
A^{\text{in}}_{0\omega} &=& i e^{\pi \beta} A^{\text{out}}_{0\omega} . \label{Ain-m0}
\end{eqnarray}
The `up' mode is
\begin{eqnarray}
\uup_{0\omega}(r) = \frac{\left(2 i \pi \omega r\right)^{1/2}}{2 \sinh(\pi \beta)} && \left[e^{\pi \beta} I_{+i\beta}(i\omega r f^{1/2})  \right. \nn \\ && \left. - e^{-\pi \beta}I_{-i\beta}(i \omega r f^{1/2})  \right].  \label{uup-m0}
\end{eqnarray}
From Eq.~(\ref{Ain-m0}) we infer that $A^{\text{in}}_{0\omega}$ has no zeros in the complex-$\omega$ plane. This leads us to conclude that there are \emph{no} quasinormal modes for $m=0$.  The solutions for higher modes $m \neq 0$ also have closed form expressions, but in terms of the lesser-known confluent Heun functions (see e.g.~\cite{Fiziev-2010}). 

 \subsubsection{Symmetries of the spectrum}
Through our time-domain simulations we found that QN ringing is a key feature in the response of higher modes ($m \neq 0$). Although only one QN mode frequency is apparent in the time-domain data of Fig.~\ref{fig:timedomain-ringing}, for each mode $m \neq 0 $ there is in fact an infinite number of damped overtones present in the QN spectrum, labeled by $n = 0, 1, \ldots, \infty$. For a given $m, n$, there is a pair of modes: one co-rotating ($+$) and one counter-rotating ($-$) with the circulating flow, with frequencies $\omega_{mn}^\pm(C)$. The spectrum has the following symmetries:
\beq
\omega^\pm_{m,n}(C) = -\omega^{\pm\ast}_{-m,n}(C)   \label{qnsym-1}
\eeq
and
\beq
\omega^\pm_{m,n}(C) = \omega^{\mp}_{-m,n}(-C) , \label{qnsym-2}
\eeq
where ${}^\ast$ denotes complex-conjugation. 
Note that (for $C > 0$) the co-rotating mode oscillates and decays faster than the counter-rotating mode, i.e.~$|\text{Re} (\omega_{mn}^+)| \ge |\text{Re} (\omega_{mn}^-)|$ and $|\text{Im} (\omega_{mn}^+)| \ge |\text{Im} (\omega_{mn}^-)|$ (with equality if $C = 0$), and the frequencies scale linearly with $1/r_h$. 

  \subsection{The continued-fraction method\label{sec:ctd-frac}}
The \emph{continued fraction method} is a fast and accurate numerical method for determining QN frequencies. It was first applied to determine the QN spectrum of black holes in Ref. \cite{Leaver-1985}, and was adapted to the draining bathtub case in Ref. \cite{Cardoso-Lemos-Yoshida}. Here we briefly recap the method giving some results in Sec.~\ref{sec:validation}.

The starting point is an ansatz for the QN wavefunction,
\beq
\psi_{m\omega}(r) = e^{i \omega r} \left( \frac{r-1}{r+1} \right)^{- i \omtil / 2} \sum_{k=0}^\infty a_k \left( 1 - r^{-1} \right)^k .  \label{cf-ansatz}
\eeq
It was  shown in Ref. \cite{Cardoso-Lemos-Yoshida} that the coefficients $a_k$ satisfy a four-term recurrence relation, and, by making use of Gaussian elimination, it may be reduced to a three-term relation, 
\beq
\alpha_k a_{k+1} + \beta_k a_k + \gamma_k a_{k-1} = 0. 
\eeq
Here $\alpha_k, \beta_k$ and  $\gamma_k$, are complex coefficients that depend upon the frequency $\omega$ and also upon $m, \B$ and $\A$. For Eq.~(\ref{cf-ansatz}) to represent a valid QN mode solution, the sum $\sum_{k} a_k$ must converge to a finite value. This condition is equivalent to the following continued-fraction condition:
\beq
\beta_0 - \frac{\alpha_0 \gamma_1}{\beta_1 -} \frac{\alpha_1 \gamma_2}{\beta_2 -} \frac{\alpha_2 \gamma_3}{\beta_3 -} \ldots = 0.  \label{ctdfrac-condition}
\eeq
Finding the roots of Eq. (\ref{ctdfrac-condition}) in the complex-frequency domain is a straightforward task for a numerical minimization algorithm. High numerical accuracy for the QN frequencies may be obtained. 

\begin{figure*}
   \includegraphics[width=8.4cm]{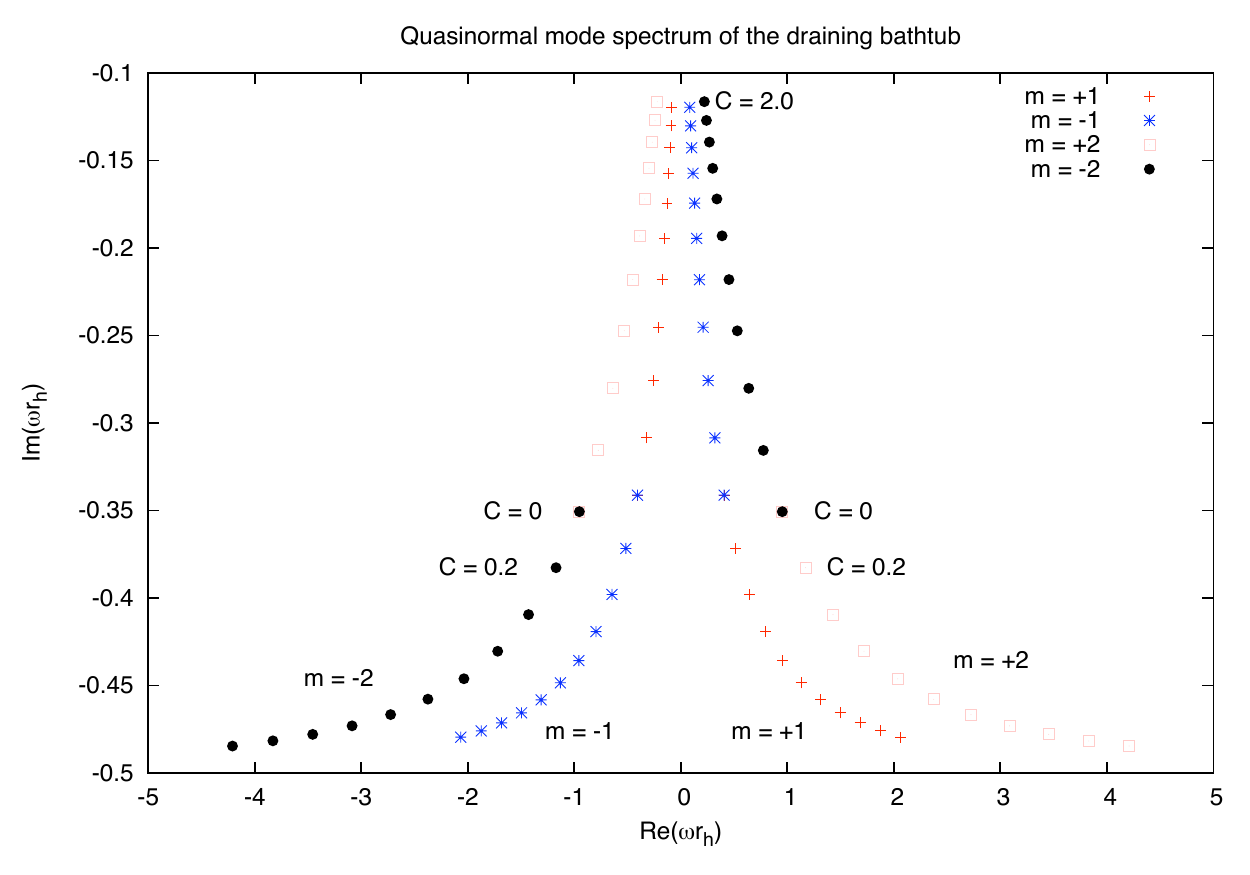}
   \includegraphics[width=8.4cm]{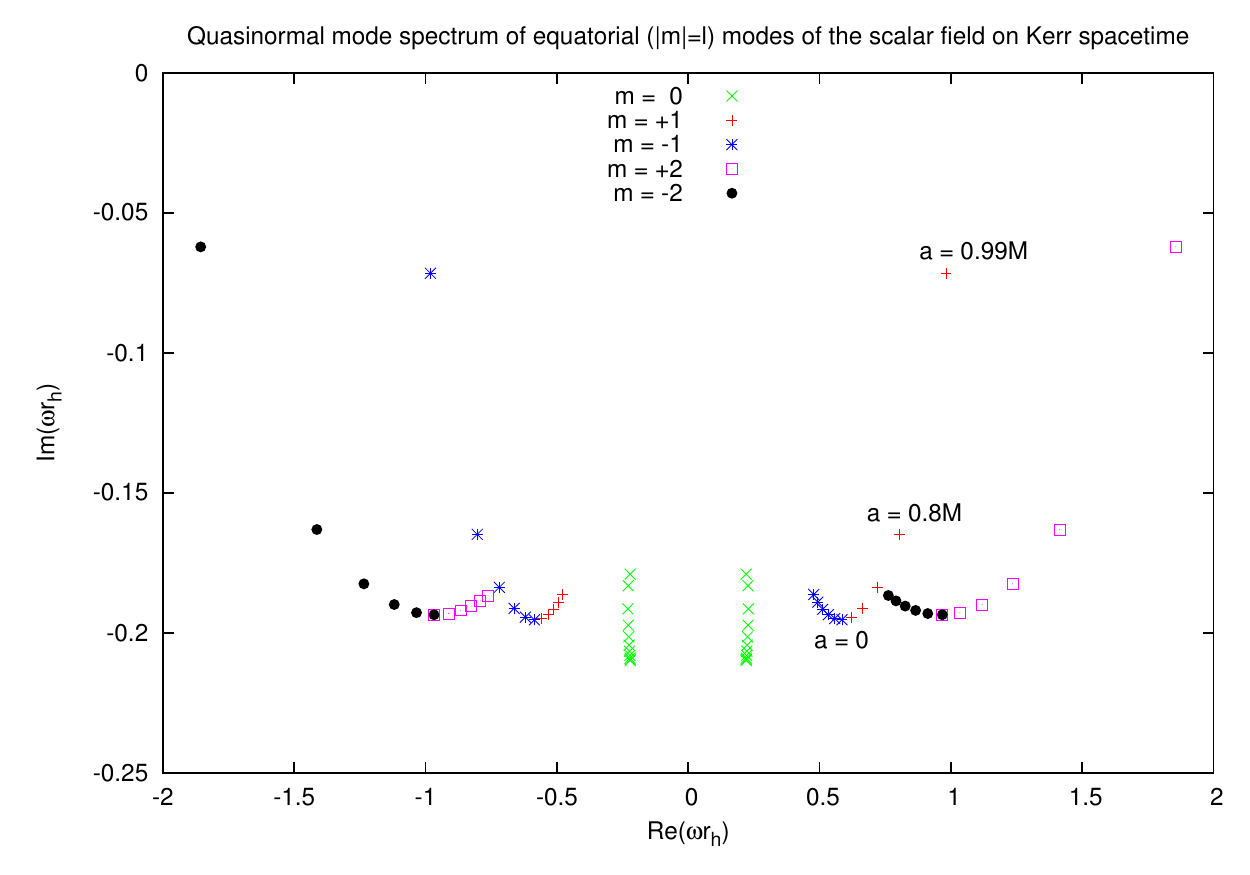}
    \caption{\emph{Left:} QN mode spectrum of the draining bathtub. \emph{Right:} QN mode spectrum of the equatorial modes of the scalar field on Kerr spacetime. 
The left plot shows the fundamental ($n=0$) quasinormal mode frequencies $\omega_{mn}^\pm$ of modes $m = \pm1$ and $\pm2$, for a range of circulation rates, $C=0, 0.2, 0.4, \ldots 2.0$. For each mode $m \neq 0$ there are two QN frequencies $\omega_{m}^\pm$. In the non-circulating case ($C=0$), the spectrum has the symmetry $\omega_{m}^{+} = -\omega_{m}^{-\ast}$. Circulation creates a difference between co-rotating ($\omega^+_{m}$) and counter-rotating ($\omega^-_{m}$) modes: Counter-rotating modes oscillate and decay more slowly than co-rotating modes ($|\text{Re} (\omega_m^+)| > |\text{Re} (\omega_m^-)|$ and $|\text{Im} (\omega_m^+)| > |\text{Im} (\omega_m^-)|$). The spectrum obeys the symmetries of Eqs.~(\ref{qnsym-1}) and (\ref{qnsym-2}).
The right plot shows the fundamental ($n=0$) quasinormal mode frequencies $\omega_{lmn}^{\pm} r_h$ of equatorial ($|m|=l$) modes $m = 0, \pm1$ and $\pm2$ of the scalar field on the Kerr spacetime, for a range of the black hole rotation rates $a / M = 0, 0.2, 0.4, 0.6, 0.8, 0.99$. Co-rotating modes oscillate more rapidly than counter-rotating modes, $|\text{Re} (\omega_{lmn}^{+})| \ge |\text{Re} (\omega_{lmn}^{-})|$. In general, both co- and counter-rotating modes become less damped as the rotation rate increases.
}
    \label{fig:qn-spectrum}
  \end{figure*}

In the left plot of Fig.~\ref{fig:qn-spectrum} we show the dependence of the `fundamental' ($n=0$) mode on $m$ and on the rotation rate $C$ of the acoustic hole. The spectrum exhibits the symmetries (\ref{qnsym-1}) and (\ref{qnsym-2}). In the non-rotating case ($C=0$) the two modes $\omega_{mn}^{+}$ and $\omega_{mn}^{-}$ are symmetric under reflection in the imaginary axis. The introduction of rotation ($C \neq 0$) breaks this symmetry. The co-rotating ($+$) mode increases the magnitude of its real and imaginary parts, whereas the counter-rotating mode ($-$) moves in the opposite way. In the limit of very large $C$, the imaginary part of the least-damped ($n=0$) counter-rotating mode asymptotes to zero, whereas the imaginary part of the co-rotating mode increases without bound.

Somewhat different behavior is observed in the spectrum of equatorial modes ($m=l$) of the scalar field of the Kerr black hole (BH), illustrated in the right plot of Fig.~\ref{fig:qn-spectrum}. In the Kerr case, the relevant rotation parameter is $a = J / M$ (rather than $C$), where $J$ and $M$ are the angular momentum and mass of the BH, respectively. As in the DBT case, the co-rotating ($+$) modes oscillate more rapidly than the counter-rotating modes ($-$), $|\text{Re} (\omega_{lmn}^{+})| \ge |\text{Re} (\omega_{lmn}^{-})|$. However, unlike the DBT, the imaginary part of both modes ($\pm$) is found to decrease in magnitude as the rotation rate increases. 

\subsection{Geometric interpretation\label{subsec:geometric}}
This behaviour may be understood through an approximate formula for the QN frequencies \cite{Goebel, Mashhoon},
\beq
\omega_{mn}^{\pm} \approx \Omega^{\pm} m - i \Lambda^{\pm} (n+1/2).
\eeq
Here, $\Omega^{\pm}$ is the orbital frequency of the prograde ($+$) or retrograde ($-$) null orbit, and $\Lambda^{\pm}$ is the corresponding \emph{Lyapunov exponent} \cite{Lyapunov1, Lyapunov3}. In the case of the DBT, we have
\beq
\Omega^{\pm} = 1 / \Lc^\pm  , \quad \quad \Lambda^{\pm} = 1 / r_c^{\pm} ,
\eeq
where $\Lc^\pm$ and $r_c^\pm$ are given in Eqs.~(\ref{lc}) and (\ref{rc}), respectively. Since $r_c^{-}$ and $\Lc^-$ increase in magnitude with $C$, the magnitude of the real and imaginary parts of $\omega_{mn}^-$ decreases with $C$ (and the opposite is true for $r_c^+$, $\Lc^+$ and $\omega_{mn}^+$). By comparison, for the equatorial modes of the Kerr BH one has instead
\beq
\Omega^{\pm} = 1/b_c^\pm, \quad \quad \Lambda^\pm = \frac{1 - 2 a / b_c^{\pm}}{\sqrt{(b_c^{\pm})^2 - a^2 }} ,
\eeq
where
\begin{eqnarray}
r_c^{\pm} &=& 2 M \left[ 1 + \cos \left( \frac{2}{3} \cos^{-1}(\mp a / M) \right) \right], \\
b_c^\pm   &=& \pm 3 \sqrt{M r_c^\pm} - a ,
\end{eqnarray}
for the equatorial orbit \cite{Mashhoon}. Expanding $\Lambda^{\pm}$ gives
\beq
\Lambda^{\pm} = (27)^{-1/2} \left[ 1 - \frac{2a^2}{27} \mp \frac{10\sqrt{3}}{243} a^3 + \mathcal{O}(a^4) \right].
\eeq
Clearly, for both co- and counter-rotating equatorial orbits on Kerr, $\Lambda^{\pm}$ decreases with $|a|$. This `explains' the observation in the right plot of Fig.~\ref{fig:qn-spectrum} that the damping decreases with $|a|$. 

In Ref.~\cite{Dolan}, a more accurate approximation for the QN modes of the Kerr BH was found using a `geodesic expansion method', which builds upon the understanding of the properties of the null orbits. Let us now apply this method in the DBT case. 
   
  \subsection{The geodesic expansion method\label{sec:expansion}}


First let us rewrite the function $u\left(r\right)$ appearing in Eq.~(\ref{radeq}) using the following ansatz \cite{Dolan-Ottewill, Dolan}:
\begin{equation}
u_{m} \left(r\right)=\chi\left(r\right)\exp\left[i \int^{r_\ast} \alpha(r') \, dr_\ast'\right],
\label{mode1}
\end{equation}
where
\begin{equation}
\alpha=\left(\omega-\frac{\B m}{r^2}\right)\left(1-\frac{\B l_c}{r^2}\right)^{-1}\left(1-\frac{r_c^{2}}{r^2}\right).
\label{alp-def}
\end{equation}
Note that here $r_c = r_c^{\pm}$ and $\Lc = \Lc^{\pm}$, i.e.~representing either the co- or counter-rotating cases, as defined in Eqs. (\ref{rc}) and (\ref{lc}), respectively;
we drop the $\pm$ superscript here for clarity. 
Substituting ansatz (\ref{mode1}) into the radial equation (\ref{radeq}) leads to an equation for the function $\chi \left(r\right)$, namely
\begin{eqnarray}
f\chi''+\left(f'+2i\alpha\right)\chi'+\left(i\alpha'+\frac{\Lc^2\left(\omega-\frac{\B m}{r^2}\right)^2}{r^2\left(1-\frac{\B \Lc}{r^2}\right)^2}\right.\nonumber\\
\left.-\frac{\left(m^2-\frac{1}{4}\right)}{r^2}-\frac{5\A^2}{4r^4}\right)\chi=0 ,
\label{eqnlam}
\end{eqnarray}
where $'$ denotes differentiation with respect to $r$.
Now, to seek the QN frequencies $\omega_{m n}$ and radial wave functions $\chi \left(r\right)$ we introduce an expansion in terms of inverse powers of $m$, namely
\begin{equation}
\Lc \omega_{mn} = \sum^{\infty}_{q=-1} {m^{-q} \varpi_q^{(n)}},
\label{wn}
\end{equation}
\begin{eqnarray}
\chi= \left[
\xi^n +\sum_{i=0}^{n} \sum_{j=0}^{\infty} a_{ij}^{(n)} m^{-j} \xi^{n-i}
\right] \nonumber\\
\times \prod^{\infty}_{q=0}{ \exp \left( m^{-q} S^{(n)}_q(r) \right)},
\label{Y} \nonumber\\
\end{eqnarray}
where $\xi \equiv 1- r_c^{\pm} / r $.
Here $\varpi_q^{(n)}$ and $a_{ij}^{(n)}$ are dimensionless coefficients, and $S^{(n)}_q(r)$ are smooth radial functions. To determine these unknowns, we impose a condition of regularity upon the solution at $r = r_c$.


Let us illustrate the approach by computing the QN frequencies and radial wave functions for the fundamental mode $(n=0)$ for the co-rotating case.  
Substituting the expansions in Eqs.~(\ref{wn}) and (\ref{Y}) into the radial equation~(\ref{eqnlam}), and then rewriting it in terms of the powers of $m$ associated with coefficients of the expansion, leads to the following system of equations:
\begin{widetext}
\begin{eqnarray}
{\cal  O}(m^2)&:& \hspace{0.5cm} \frac{\left(\varpi_{-1} -\frac{r_c^2 - \Lc^2/2}{r^2}\right)^2}{r^2\left(1-\frac{(r_c^2 - \Lc^2/2)}{r^2}\right)^2} - \frac{1}{r^2} = 0 \hspace{1cm} \Rightarrow \varpi_{-1} =1,  \label{Om2}
\\
\nonumber\\
{\cal  O}(m^1)&:& \hspace{0.5cm} \frac{2 r_c^2 (r^2- [r_c^2-\Lc^2/2])}{\Lc r^5} + \frac{2i}{\Lc}\left(1-\frac{[r_c^2-\Lc^2/2]}{r^2}\right)\left(1- \frac{r_c^2}{r^2} \right)S'_0
+\frac{2\varpi_0 \Lc}{r^2}=0, \label{order}
\end{eqnarray}
\end{widetext}
etc., where the superscript $n$ has been suppressed for simplicity. Equation (\ref{Om2}) is identically satisfied through the choice $\varpi_{-1} =1$. To solve Eq. (\ref{order}) we impose that $S_0(r)$ is regular at $r = r_c$, which leads to $\varpi_{0} = -i \Lc / 2 r_c $
and to the first-order differential equation
 \begin{eqnarray}
(r-r_c) (r+r_c) \frac{dS^{(0)}_0}{dr} = - \frac{r_c^2}{r} - \frac{l_c^2}{2 r_c} \left(1 - \frac{l_c C}{r^2} \right)^{-1} . \label{S0prime} 
\end{eqnarray} 
The radial function $S_0$ may be obtained by integrating this equation. In subsequent equations at orders $\mathcal{O}(m^{0})$ and higher, the second derivative $S_0''$ appears, which may be obtained by differentiating (\ref{S0prime}). 
The higher-order equations are solved in a similar way: taking the equation at $\mathcal{O}(m^{1-k})$ we first impose the continuity condition at $r=r_c$ to obtain $\varpi_{k}$, and next solve to obtain $S_k'$. 


Using a symbolic algebra package (e.g.~Maple or Mathematica), this procedure may be automated, and the expansion may be taken to higher orders. We have computed the expansion of the QN frequency up to the order ${\cal O}(m^{-9})$. Below, we quote the expansion of the frequency of a general overtone $n$ ($= N - 1/2$) up to order ${\cal O}(m^{-4})$:
\begin{widetext}
\begin{eqnarray}
\Lc \, \omega_{mn}=m-\frac{i\Lc}{r_c}N+m^{-1}\left[ \frac{1}{128}\frac{\Lc^2}{r_c^4}\left(5\Lc^2-16r_c^2\right)-\frac{3}{32}\frac{\Lc^4}{r_c^4}N^2\right]
+m^{-2}i\left[\frac{1}{4096}\Lc^3\left(5\Lc^4-144\Lc^2r_c^2+384r_c^4\right)\frac{1}{r_c^7}N\right. \nonumber\\
\left.
+\frac{1}{1024}\Lc^5\left(23\Lc^2-80r_c^2\right)\frac{1}{r_c^7}N^3\right]
+m^{-3}\frac{1}{r_c^{10}}\left[ -\frac{1}{1048576}\Lc^4\left(64640\Lc^2r_c^4-21040r_c^2\Lc^4+2125\Lc^6-57344r_c^6\right)\right.\nonumber\\
\nonumber\\
+\left.\frac{1}{131072}\Lc^4\left(3456\Lc^2r_c^4-976r_c^2\Lc^4+75\Lc^6-4096r_c^6\right)N^2+\frac{5}{65536}\Lc^6\left(896r_c^4-592\Lc^2r_c^2+91\Lc^4\right)N^4\right] + {\cal O}(m^{-4}) . \nonumber\\
\label{freqex}
\end{eqnarray}
\end{widetext}
Note that, for a given $m$, this expression yields two QN frequencies: $\omega_{mn}^+$ obtained using co-rotating geodesic parameters ($\Lc^+, r_c^+$) and $\omega_{mn}^-$ obtained using counter-rotating geodesic parameters ($\Lc^-, r_c^-$). The frequency expansion exhibits the symmetries (\ref{qnsym-1}) and (\ref{qnsym-2}), as may be confirmed with the aid of the relations $r_c^{\pm}(C) = r_c^{\mp}(-C)$ and $\Lc^{\pm}(C) = -\Lc^{\mp}(-C)$. 
  
 \subsection{Validation\label{sec:validation}}
\begin{figure}
  \includegraphics[width=8.2cm]{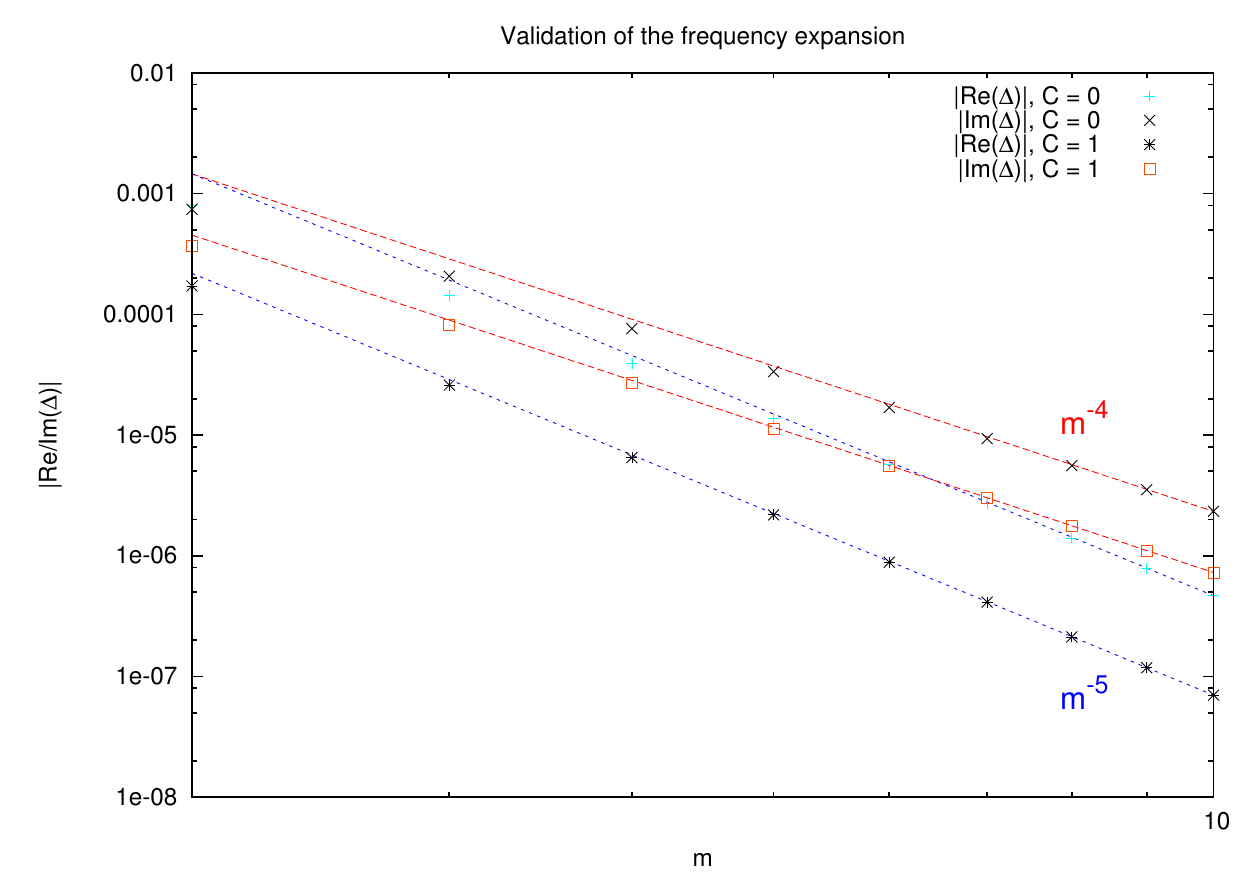}
  \caption{Validation of the frequency expansion, Eq.~(\ref{freqex}). The log-scaled plot shows the real and imaginary parts of $\Delta$ as a function of mode number $m$, for the cases $\B = 0$ and $\B = 1$ (and $\A = 1$). Here $\Delta \equiv \omega_{\text{CF}} - \omega_{\text{exp}}$, where $\omega_{\text{CF}}$ is the QN frequency obtained via the continued fraction method and $\omega_{\text{exp}}$ is the frequency estimate given by Eq.~(\ref{freqex}). The plot shows that the imaginary (real) part of $\Delta$ scales as $m^{-4}$ ($m^{-5}$), as predicted by Eq.~(\ref{freqex}).}
  \label{fig:validation}
 \end{figure}

 In Table \ref{table1} we compare the QN frequencies found via Eq.~(\ref{freqex}) with numerically-accurate values obtained via the continued-fraction method of Ref. \cite{Cardoso-Lemos-Yoshida}, and re-obtained here in Sec.~\ref{sec:ctd-frac}. At large $m$, we find excellent agreement. At small $m$, the approximation is not so good. For instance, for $m=1$ the most accurate estimate was found by truncating the series at $\mathcal{O}(m^{-3})$, suggesting that Eq. (\ref{freqex}) is in fact an asymptotic series. In Fig.~\ref{fig:validation} we plot the difference between continued-fraction and geodesic-expansion results, to confirm that Eq.~(\ref{freqex}) is indeed valid to the stated order in the large-$|m|$ regime.  
 
\begin{table*}
\caption{QN frequencies of the fundamental mode ($n=0$) for $m=1\ldots5$. The second column gives the frequencies determined via the continued-fraction method of Sec.~\ref{sec:ctd-frac}. The third column gives the frequencies estimated from the expansion method, Eq.~(\ref{freqex}). The numeral in parentheses indicates the absolute error in the last displayed digit, where the error in the expansion method was estimated from the magnitude of the final terms in the series, Eq.~(\ref{freqex}). Note that the value marked with an asterisk (${}^*$) was obtained by truncating the series at $\mathcal{O}(m^{-3})$.}
\begin{center}

\begin{tabular}{l | l | l}
\hline \hline
\multicolumn{3}{l}{$\A = 1$, $\B = 0$, symmetric modes ($\pm$)} \\
\hline \hline
$m$ & Continued-Fraction & Expansion Method \\
\hline
$1$ & \, $\pm 0.4068326196672(2)  -0.341236118125(2)i$ & $\pm 0.40(1){}^* - 0.345(8) i{}^* $  \\
$2$ & \, $\pm 0.95272808772474(6) -0.3507394957317(2) i$ & $\pm 0.9524(9) - 0.3511(2) i$  \\
$3$ & \, $\pm 1.46854069662523(5) -0.3524255329360(2) i$ & $\pm 1.46854(6) - 0.352444(8) i$  \\
$4$ & \, $\pm 1.9764527143560(1) -0.3529594212624(2) i$ & $\pm 1.976453(8) - 0.352961(1) i$  \\
$5$ & \, $\pm 2.4811874975616(1) -0.35318833783278(6) i$ & $\pm 2.481188(2) - 0.3531886(1) i$  \\
\hline
\hline
\end{tabular}

\begin{tabular}{l | l | l}
\hline
\hline
\multicolumn{3}{l}{$\A = 1$, $\B = 1$, counter-rotating modes ($-$)} \\
\hline
\hline
$m$ & Continued-Fraction & Expansion Method \\
\hline
\hline
$1$ & \, $-0.1490148506555(1) - 0.1945123420127(2) i$ & $- 0.14(1) - 0.200(8) i$  \\
$2$ & \, $-0.3875750711220(2)  - 0.1930307208974(1) i$ & $- 0.38753(6) - 0.19296(9) i$  \\
$3$ & \, $-0.6040167016214(2) - 0.19220554998274(1) i$ & $- 0.604014(4) - 0.192204(3) i$  \\
$4$ & \, $-0.8155812868761(1) - 0.1918516723182(3) i$ & $- 0.8155811(5) - 0.1918515(3) i$  \\
$5$ & \, $-1.0253078467593(1) - 0.1916753960108(2) i$ & $- 1.0253078(1) - 0.19167538(6) i$  \\
\hline
\hline
\end{tabular}

\begin{tabular}{l | l | l}
\hline \hline
\multicolumn{3}{l}{$\A = 1$, $\B = 1$, co-rotating modes ($+$)} \\
\hline \hline
$m$ & Continued-Fraction & Expansion Method \\
\hline
$1$ & \, $+1.13097639081141(2) - 0.4485344193663(3) i$ & $+ 1.134(9) - 0.445(7) i$  \\
$2$ & \, $+2.3742286767072(2) - 0.45797306346650(6) i$ & $+ 2.37421(3) - 0.45791(2) i$  \\
$3$ & \, $+3.5943282234311(3)  - 0.4600949990202(3) i$ & $+ 3.594327(2)  - 0.460093(2) i$  \\
$4$ & \, $+4.8080846946733(2) - 0.46088327915183(4) i$ & $+ 4.8080845(3) - 0.4608831(2) i$  \\
$5$ & \, $+6.0192217565670(3) - 0.4612576561145(2) i$ & $+ 6.01922173(6) - 0.46125763(3) i$  \\
\hline
\hline
\end{tabular}

\label{table1}
\end{center} 
\end{table*}

\section{Regge pole resonances\label{RPs}}
Regge poles (RPs) are closely related to QN modes \cite{newton}.
Both types of resonance are associated with the zeros of $\Ain$. Whereas QN modes occur at real values of $m$ (and complex $\omega$), the Regge poles (RPs) occur for real values of $\omega$ (and complex $m$). That is, $m_{\omega n}$ is a Regge pole angular momentum if
\beq
\Ain(\omega, m_{\omega n}) = 0.
\eeq

\subsection{Methods}
Given the close relationship between QN modes and RPs, it is no surprise to find that methods used for QN modes can be easily adapted to locate RPs \cite{Glampedakis-Andersson, Decanini-Folacci, Decanini-Folacci-Jensen, DFR-rp}. For instance, the continued-fraction method (Sec.~\ref{sec:ctd-frac}) may be used without modification, if $m$ is allowed to become complex. Furthermore, the geodesic expansion method (Sec.~\ref{sec:expansion}) can be easily modified to find RPs \cite{Dolan-Ottewill}. Either one may repeat the arguments of Sec.~\ref{sec:expansion}, or one may take the following simple steps: (i) assume that the Regge poles have the expansion of the form
\begin{equation}
m_{\omega n}  = \zeta_{-1} \omega + \zeta_0 + \zeta_{1} \omega^{-1} + \ldots ,
\end{equation}
(ii) substitute the expansion for QN frequencies into the above equation, and (iii) solve order-by-order in $m$ to find the expansion coefficients $\zeta_{k}$. 
We find
\begin{widetext}
\begin{eqnarray}
m_{\omega n} &=& \Lc \omega+\frac{i \Lc}{r_c}N+\left[ \frac{3 \Lc^3 N^2}{32 r_c^4} 
+ \frac{\Lc \left( 16r_c^2-5\Lc^2\right)}{128r_c^4} \right] \omega^{-1}
- i N \left[ \frac{\Lc}{4096r_c^7}\left( -304r_c^2 \Lc^2+5 \Lc^4+896r_c^4\right) \right. \nonumber\\
\nonumber\\
&&\left.  + \frac{\Lc^3}{1024r_c^7} \left( 16r_c^2+23\Lc^2\right) N^2 \right] \omega^{-2}+
\left[ \frac{\Lc \left( - 73728 r_c^6 + 74880  r_c^4 \Lc^2 - 22640 r_c^2 \Lc^4 + 2125 \Lc^6  \right)}{1048576 r_c^{10}}
\right. \nonumber \\
&& \left. 
- \frac{\Lc \left(36864r_c^6 - 7808r_c^4 \Lc^2 - 1616 r_c^2 \Lc^4 + 75\Lc^6 \right)}{131072 r_c^{10}}  N^2
-\frac{\Lc^3 \left(384r_c^4 + 560 r_c^2 \Lc^2 + 455 \Lc^4 \right)}{65536 r_c^{10}} N^4 \right] \omega^{-3} + 
{\cal O}(\omega^{-4}) ,
\label{momex}
\end{eqnarray}
\end{widetext}
where $N=n+1/2$. Again, as for Eq.~(\ref{freqex}), we note that this expression yields two separate values, $m_{\omega n}^{\pm}$, obtained by using either co-rotating or counter-rotating geodesic parameters ($\Lc^\pm, r_c^\pm$). We note that it is straightforward to take the expansion to much higher orders if desired. Finally, it is relatively straightforward to find Regge poles by direct integration methods, because (unlike quasinormal modes), RP modes are not divergent in the limits $r_\ast \rightarrow \pm \infty$.
  
  \subsection{Spectrum}
The Regge pole spectrum has the following symmetries
\beq
m_{\omega,n}^{\pm}(C) = -m_{-\omega,n}^{\pm\ast}(C)  \label{rpsym-1}
\eeq
and
\beq
m_{\omega,n}^{\pm}(C) = - m_{\omega, n}^{\mp}(-C).  \label{rpsym-2}
\eeq
 
\begin{figure}
 \includegraphics[width=8.2cm]{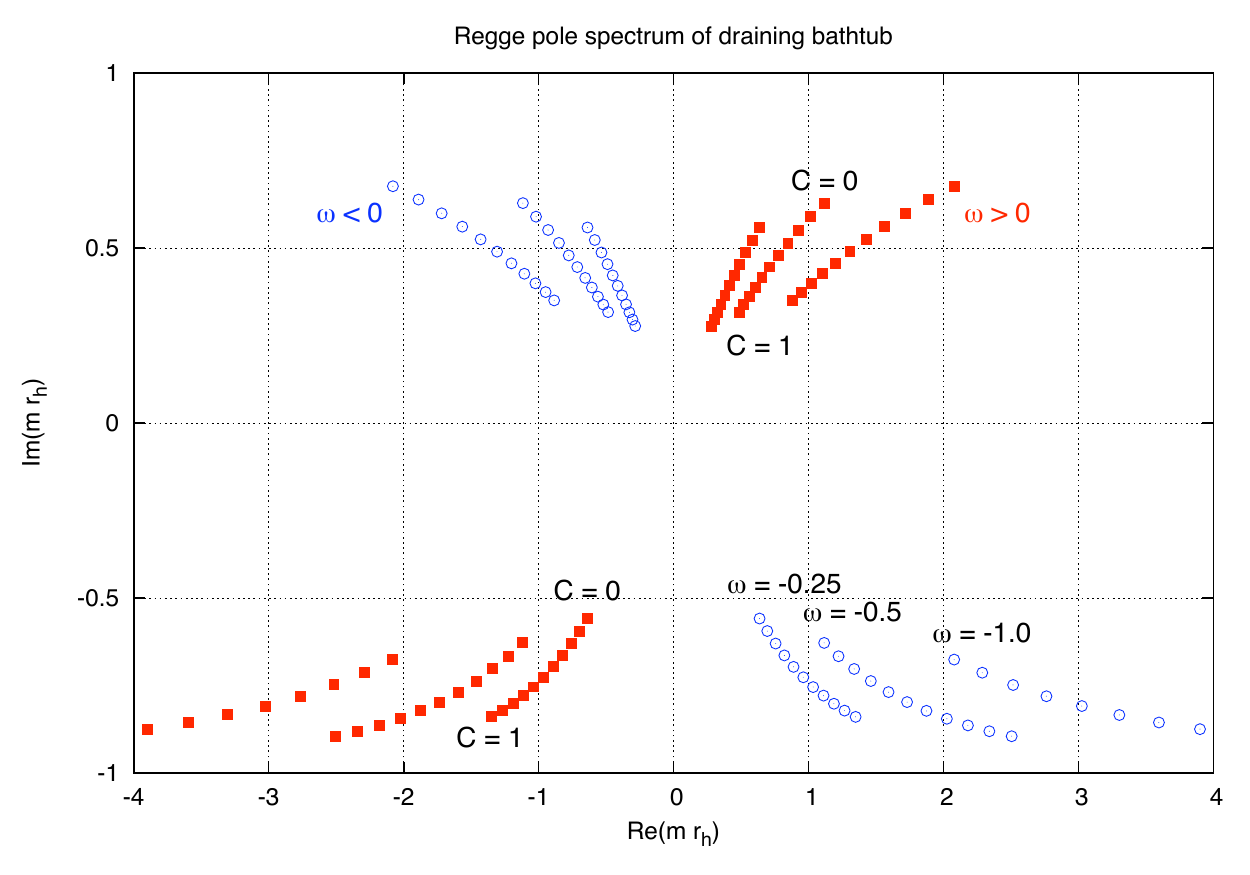}
  \caption{Regge pole spectrum of the draining bathtub. For three frequencies $|\omega| = 0.25$, $0.5$, and $1.0$, we plot points for rotation rates $\B/r_h = 0, 0.1, 0.2, \ldots, 1$, showing the fundamental ($n=0$) Regge pole values $m_{\omega n}^{\pm} r_h$ in the complex-$m$ plane. Co-rotating and counter-rotating modes are shown; they exhibit the symmetries (\ref{rpsym-1}) and (\ref{rpsym-2}). For $C > 0$, the co-rotating (counter-rotating) modes lie in the upper (lower) half plane. }
  \label{fig:rp-spectrum}
 \end{figure}

The Regge pole spectrum in the complex-$m$ plane is illustrated in Fig.~\ref{fig:rp-spectrum}. For $\omega > 0$, the co-rotating $(+)$ modes lie in the first quadrant, and the counter-rotating ($-$) modes lie in the fourth quadrant. In the non-rotating case, the spectrum is symmetric, $m_{\omega n}^+ = - m_{\omega n}^-$. For $C > 0$, the real and imaginary parts of $m_{\omega n}^+$ decrease with $C$, whereas the real and imaginary parts of $-m_{\omega n}^-$ increase with $C$, as expected from Eq. (\ref{momex}). 
 
  \section{Applications of Complex Angular Momentum (CAM) method\label{sec:CAM}}
In this section we show that Regge poles play an important role in both absorption and scattering processes; their role is revealed through application of the so-called Complex Angular Momentum (CAM) method. The CAM method makes a link between certain oscillations in cross sections (see below) and the poles of the scattering matrix $S(m)$ in the complex-$m$ plane, i.e.~the Regge poles. The CAM method was successfully applied to scattering by a Schwarzschild black hole in \cite{Andersson-1994}, and was recently extended to treat absorption \cite{DEFF, DFR, DFR2}. We believe this represents its first application to a rotating spacetime.
  
  \subsection{Absorption Cross Section\label{subsec:abs-csec}}
\begin{figure*}
\includegraphics[width=8.2cm]{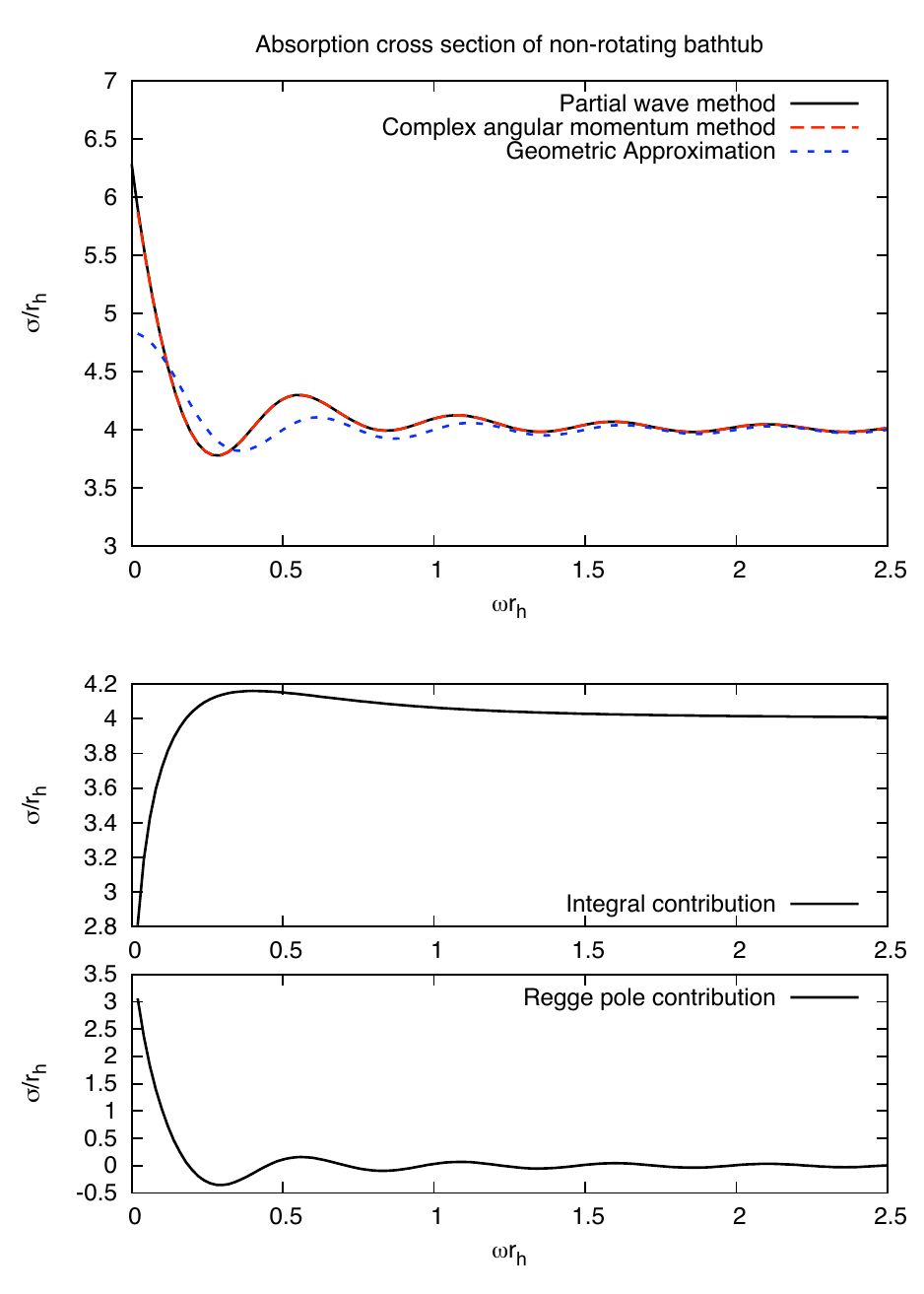}
\includegraphics[width=8.2cm]{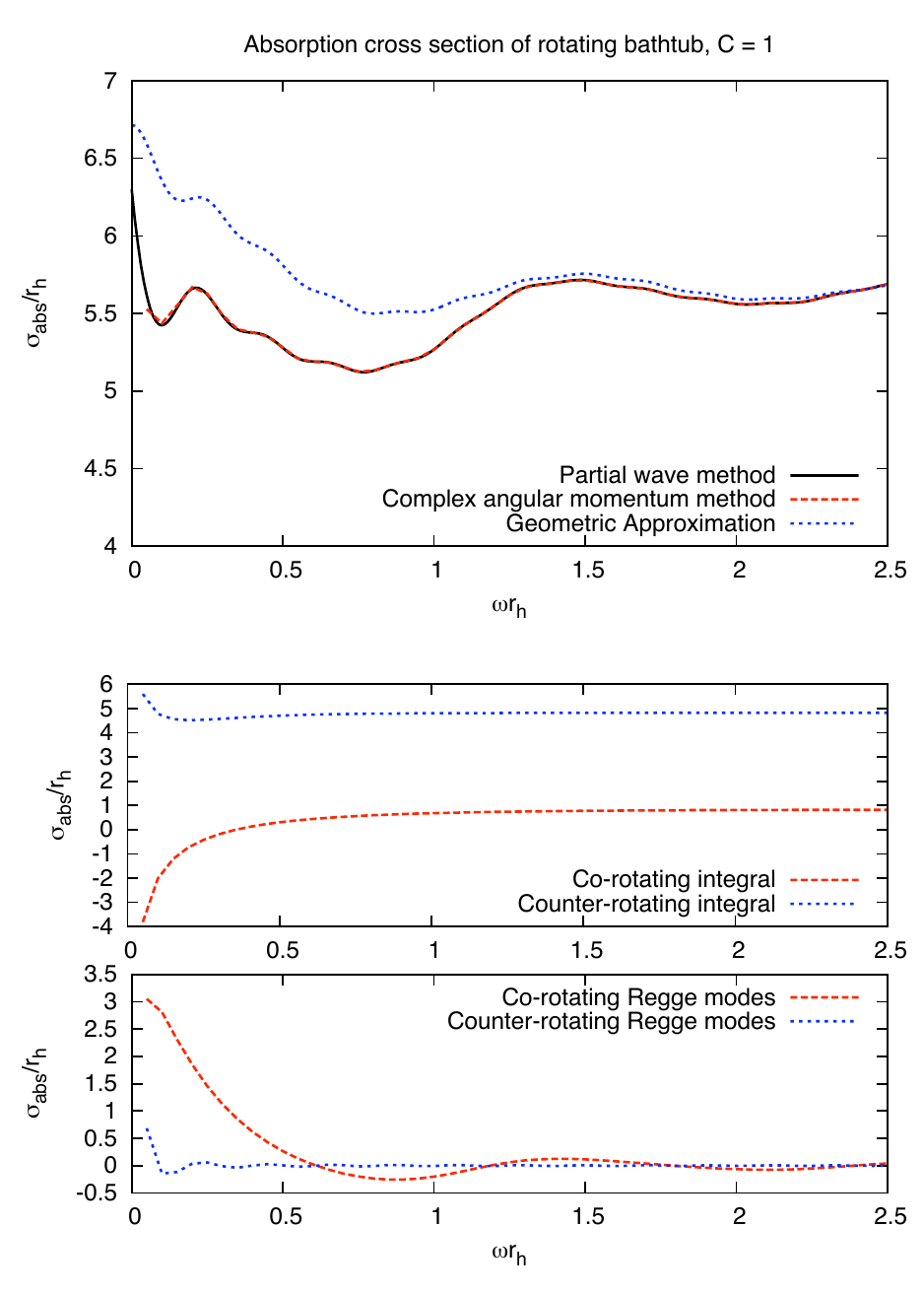}
\caption{Absorption cross section as a function of the frequency, for $C = 0$ (left plots) and $C/r_h = 1$ (right plots). In the upper plots, the solid line shows the absorption cross section $\sigabs$ computed from the partial wave sum \cite{ODC}, the dashed line shows $\sigabs$ computed numerically from the complex angular momentum expression [Eq.~(\ref{sig-cam})]. Note that these two lines lie very close together. The dotted line shows the simple geometric approximation [Eq.~(\ref{abs-geo})], which is valid in the high-frequency regime. The lower plots show the separate contributions from Regge pole sums and the co-rotating and counter-rotating integrals [Eq.~(\ref{sigabs-decomp})].}
\label{fig-abs}
\end{figure*}

The absorption cross section $\sigabs$ of the draining bathtub was studied in Ref.~\cite{ODC}. It may be obtained from a sum over partial wave contributions, via
\beq
\sigabs = \frac{1}{\omega} \sum_{m=-\infty}^{\infty} \left( 1 - \left| \frac{\Aout}{\Ain} \right|^2 \right).
\eeq
At high frequencies, the absorption cross section is seen to approach the `geometric' capture cross section, $\sigabs^{\text{(geo)}} = |b_c^+| + |b_c^-| = 4 \sqrt{\A^2 + \B^2}$. In the non-rotating case, $C=0$, $\sigabs$ displays regular damped oscillations with increasing frequency. With increasing $C$, these oscillations become less regular (cf.~Fig.~\ref{fig-abs}). 

Recent papers \cite{DEFF, DFR, DFR2} on the universality of high-frequency absorption cross sections for \emph{spherically-symmetric} black holes use CAM methods \cite{Andersson-Thylwe-1994, Andersson-1994, newton, Nussenzveig} to show how oscillations in the absorption cross section (with frequency $\omega$) are directly related to the Regge pole spectrum. 

\begin{figure}
\includegraphics[width=8cm]{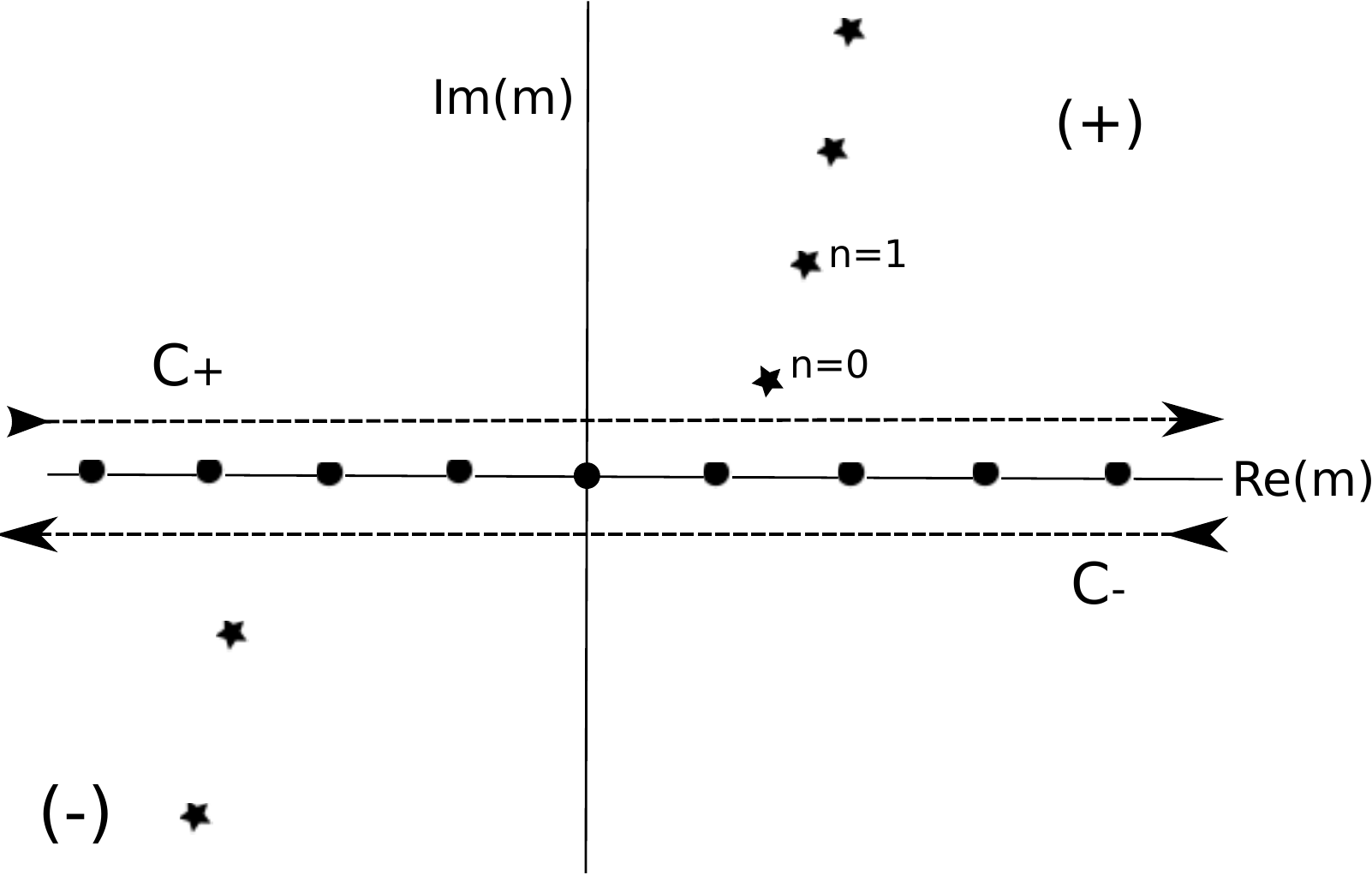}
\caption{Contour integration in the complex-$m$ plane. The circles show the positions of the poles of $1/\sin(m\pi)$, at integer values of $m$. The stars illustrate the positions of the poles (for $\omega > 0$, $C > 0$) of $S(m)$, i.e.~the Regge poles, for overtones $n = 0 \ldots \infty$.  The contour $\mathcal{C}_+$ is closed in the upper half-plane, and the contour $\mathcal{C}_-$ is closed in the lower half-plane. 
}
\label{fig-watson}
\end{figure}

To apply CAM methods, we must employ a suitable analytic continuation into the complex-$m$ plane. Let us therefore define
\beq
\sigabs = \frac{1}{\omega} \sum_{m=-\infty}^{\infty} \Gamma(m),
\eeq
where
\beq
\Gamma(m) = 1 -  \frac{\Aout (A^{\text{out}}_{m^\ast \omega})^\ast}{\Ain (A^{\text{in}}_{m^\ast \omega})^\ast}.
\eeq
A series such as the one above can be expressed as a contour integral using the Watson transformation, i.e.
\beq
\sigabs = \frac{i}{2 \omega} \int_{\mathcal{C}} \frac{e^{i \pi m}}{\sin(\pi m)} \Gamma(m) d m ,
\eeq
where now $m$ takes complex values. Here the contour of integration $\mathcal{C}$ encloses in a clockwise sense all (and only) the poles of the integrand that lie on integer values on the real axis at $m = -\infty, \ldots, +\infty$, i.e.
\beq
\int_{\mathcal{C}}  = \int_{\mathcal{C+}} - \int_{\mathcal{C-}} , \quad \text{where} \quad \int_{\mathcal{C\pm}} = \int_{-\infty \pm ic}^{\infty \pm ic} ,
\eeq
for some small positive real value $c$. See Fig.~\ref{fig-watson} for a graphical representation of the contour. Next, we use the identity $e^{i \pi m} = e^{-i \pi m} + 2i \sin(\pi m)$ to obtain 
\begin{eqnarray}
\sigabs = \sigabs^{\text{(int)}} && + \frac{i}{2\omega} \int_{\mathcal{C}_+} \frac{e^{i \pi m} \Gamma(m)}{\sin(\pi m)} dm \nonumber \\ && - \frac{i}{2\omega} \int_{\mathcal{C}_-} \frac{e^{-i \pi m} \Gamma(m)}{\sin(\pi m)} dm ,
\label{sigabs-decomp}
\end{eqnarray}
where
\beq
\sigabs^{\text{(int)}} = \frac{1}{\omega} \int_{-\infty}^{\infty} \Gamma(m) dm .  \label{sig-int}
\eeq
The contour $\mathcal{C}_+$ may be closed in the upper half-plane, enclosing poles of $\Gamma(m)$ at $m_{\omega n}^{+}$ and $m_{\omega n}^{- \ast}$. Likewise, the contour $\mathcal{C}_-$ may be closed in the lower half-plane, enclosing the poles of $\Gamma(m)$ at $m_{\omega n}^{-}$ and $m_{\omega n}^{+ \ast}$. This leads to a sum of residues, that is,
\beq
\sigabs = \sigabs^{\text{(int)}}
 - \frac{2\pi}{\omega} \, \text{Re} \, \sum_{\pm} \sum_{n=0}^{\infty} \frac{e^{\pm i \pi m_{\omega,n}^{\pm}}} {\sin \left(\pi m_{\omega n}^{\pm} \right)} \gamma_{\omega n}^\pm.  \label{sig-cam}
\eeq
Here $\gamma_{\omega n}^\pm$ denotes the residue of $\Gamma(m)$,
\beq
\gamma_{\omega n}^\pm = \mathop{\text{Res}}_{m \rightarrow m_{\omega n}^{\pm}} \left[ \Gamma(m) \right], 
\eeq
and we have made use of the symmetry relation
\beq
\mathop{\text{Res}}_{m \rightarrow m_{\omega n}^{\pm \ast}} \left[ \Gamma(m) \right] = \left( \gamma_{\omega n}^\pm \right)^\ast .
\eeq
Equation~(\ref{sig-cam}) is an exact expression which may be computed numerically. The sum over overtones converges rapidly, due to the damping effect of the imaginary part of the Regge pole. The residues may be computed with a numerical scheme based on integration of the radial equation (\ref{radeq}), using
\beq
\gamma_{\omega n}^{\pm} = - \left( \frac{\Aout (A^{\text{out}}_{m^\ast \omega})^\ast}{\alpha^{\pm}_{\omega n} (A^{\text{in}}_{m^\ast \omega})^\ast} \right)_{m = m_{\omega n}^{\pm}},
\eeq
where $\Ain \approx \alpha^{\pm}_{\omega n} (m - m_{\omega n}^{\pm}) + \ldots$ in the vicinity of a Regge pole. The integral over frequency can also be performed in a straightforward manner. Numerical results are shown in Fig.~\ref{fig-abs}. We find a very good agreement with the results of the partial wave method of Ref.~\cite{ODC}.

To better understand the geometric meaning of Eq.~(\ref{sig-cam}) in the high-frequency limit, we may proceed by making some simplistic approximations: 
\beq
\Gamma(m) \approx \Theta( l_c^+ \omega - m) \Theta(m - l_c^- \omega),
\eeq
where $\Theta(\cdot)$ is the Heaviside step function,
and
\beq
m_{\omega, n=0}^{\pm} \approx l_c^{\pm} \omega + \frac{i l_c^{\pm}}{2 r_c^{\pm}} + \mathcal{O}(\omega^{-1}),
\eeq
and furthermore
\beq
\gamma_{\omega n}^{\pm} \approx - \frac{\eta^{\pm}}{2 \pi} + \mathcal{O}(\omega^{-1}),
\eeq
where $\eta^{\pm} = | l_c^{\pm} / r_c^{\pm} |$. Neglecting the higher overtones leads to the simple approximation 
\beq
\sigabs^{\pm} = |l_c^\pm| \left[ 1 + 4 \pi \eta^{\pm} e^{- \pi \eta^{\pm}} \sinc( 2 \pi |\Lc^\pm| \omega ) \right],
\label{abs-geo}
\eeq
where $\sigabs = \sigabs^+ + \sigabs^-$ and $\sinc(x) = \sin(x) / x$. Equation~(\ref{abs-geo}) makes it clear that the cross section approaches the geometric capture cross section in the high-frequency limit. Furthermore, the cross section exhibits damped oscillations with increasing frequency, which are due to the Regge pole contribution, which are controlled by the geometric quantities ($l_c^\pm$ and $r_c^{\pm}$). 

Note that a somewhat more accurate approximation may be obtained by using an improved approximation (cf.~Eq.~(19) in Ref.~\cite{DEFF}) for the transmission factor, 
\beq
 \Gamma(m) \approx
 \begin{cases}  
   \left[ 1 + \exp\left( - 2\pi \frac{[(\omega l_c^+)^2 - m^2]}{2 m \eta^+ } \right) \right]^{-1} , & m > 0, \\ 
  \left[ 1 + \exp\left( - 2\pi \frac{[(\omega l_c^-)^2 - m^2]}{2 m \eta^- } \right) \right]^{-1},  & m < 0.
 \end{cases}
\eeq

In Fig.~\ref{fig-abs} we compare the results of the partial-wave calculation of Ref.~\cite{ODC} with a numerical calculation based on the exact Regge pole expression, Eq.~(\ref{sig-cam}), and the `geometric' high-frequency approximation, Eq.~(\ref{abs-geo}). It is clear from the lower right plot of Fig.~\ref{fig-abs} that the co-rotating (counter-rotating) orbits are responsible for oscillations in the profile with a higher (lower) frequency and smaller (larger) amplitude. The superposition of co- and counter-rotating effects creates a rather irregular profile, in contrast with the more regular profile exhibited in the non-rotating case (left plots). 

\subsection{Scattering cross sections and `orbiting'\label{subsec:scat-csec}}

Various studies of monochromatic planar-wave scattering by black holes \cite{FHM, Anninos, Dolan-2008, electromagnetic} have shown that there may arise regular oscillations in the scattering cross section, whose angular width is inversely proportional to incident wavelength. This phenomenon is called `orbiting' \cite{Newton, Adhikari-Hussein}, or alternatively, `spiral scattering' \cite{Anninos}. In essence, such oscillations arise due to interference between pairs of rays which pass in opposite senses around the black hole. A study of scattering by the DBT \cite{DOC-AB, ODC2} has recently demonstrated that orbiting oscillations also arise in this case.

(Note that, in three-dimensional scattering, orbiting is supplemented by another effect: the `glory'. A glory is a bright spot (or ring) in the forward- or backward-scattering directions, whose width (amplitude) decreases (increases) linearly with frequency. 
Semi-classically, a glory is created by a one-parameter family of geodesics scattering into the same solid angle. This is not possible in 2D; hence the glory effect is absent in the DBT case.)

In this section, we apply the CAM method to obtain a deeper understanding of orbiting oscillations in the scattering cross section. Let us begin with the partial-wave expression for the scattering amplitude,
\beq
f_{\omega}(\phi) =  \kappa \sum_{m=-\infty}^\infty \left[ S(m) - 1 \right] e^{i m \phi} \label{fscat-def}
\eeq
where $\kappa = \left( 2 i \pi \omega \right)^{-1/2}$ and the `scattering matrix' $S(m)$ (which is scalar-valued in this case) is defined by
\beq
S(m) \equiv i e^{i \pi m} \Aout / \Ain .
\eeq
We note that, with this definition, the scattering matrix has the following symmetry:
\beq
S(-m, C) = e^{-2 i \pi m} S(m, -C).
\eeq
The differential cross section is simply $d\sigma / d\phi = |f_{\omega}|^2$. We will adopt the convention that the scattering angle is in the range $0 \le \phi < 2 \pi $ (rather than $-\pi \le \phi < \pi $).

As in Sec.~\ref{subsec:abs-csec}, the key step is to apply a Watson transformation \cite{Newton} to convert the partial wave series (\ref{fscat-def}) into a contour integral:
\beq
f_\omega(\phi) = \frac{i \kappa}{2} \int_{\mathcal{C}} \frac{\tilde{S}(m) e^{i m \phi}}{\sin(\pi m)} dm, \label{wat}
\eeq
where $\tilde{S}(m) = e^{-i \pi m}S(m)$.
 
Let us proceed by recasting (\ref{wat}) into the form
\begin{eqnarray}
&& f_\omega(\phi)  = \kappa \int_{-\infty}^{+\infty} S(m) e^{i m (\phi - 2 \pi)} dm \nonumber \\ &&+ \frac{i\kappa}{2} \int_{\mathcal{C}_+} \frac{\tilde{S}(m) e^{i m \phi}}{\sin(\pi m)} dm  - \frac{i \kappa}{2} \int_{\mathcal{C}_-}  \frac{\tilde{S}(m) e^{i m (\phi - 2\pi)}}{\sin(\pi m)} dm . \nonumber
\end{eqnarray}
In the high-frequency limit, we may neglect the first integral, because it has no stationary phase points in the range $0 < \phi < 2 \pi$. Furthermore, we may evaluate the second and third terms by closing the contours $\mathcal{C}_+$ and $\mathcal{C}_-$ in the upper and lower half-planes, respectively. Applying Cauchy's theorem leads us to
\beq
f_\omega(\phi)  \approx  - \kappa \pi \sum_{n=0}^\infty \left[ \frac{e^{i m_{\omega n}^{+} \phi}}{\sin(\pi m_{\omega n}^{+})} \tilde{s}_{\omega n}^{+} +  \frac{e^{i m_{\omega n}^- (\phi - 2\pi)}}{\sin(\pi m_{\omega n}^{-})}  \tilde{s}_{\omega n}^{-} \right] \label{cam-scat-ampl} ,
\eeq 
where the residues are defined as follows, 
\beq
\tilde{s}_{\omega n}^{\pm} \equiv \mathop{\text{Res}}_{m \rightarrow m^{\pm}_{\omega n}} \tilde{S}(m) .  \label{rp-res-def}
\eeq
The `fundamental' mode ($n = 0$) gives the dominant contribution, and the sum over overtones appears to be exponentially convergent.


\subsubsection{High-frequency approximation}
The geodesic expansion method led us to expressions for the Regge poles and corresponding wave functions, which are valid in the high-frequency regime. Now, by combining this method with standard WKB techniques, we may also obtain a low-order approximation for the residues (\ref{rp-res-def}). As the calculation is rather lengthy, we quote the key results here, and give a fuller exposition in Appendix \ref{Appendix:RP-residues}.

In the high-frequency limit, the RP wave function is approximately
\beq
u_{\omega n}(r) \approx \left( \frac{|r-r_c|}{r+r_c} \right)^n \left( \frac{r}{r+r_c} \right) \exp \left( i \omega r - i \tilde{\omega} (r_\ast - r) \right) ,
\eeq
where $\tilde{\omega} = \omega - C m^\pm_{\omega n} / r_h^2$. 
Here we have chosen the normalisation such that $u_{\omega n}(r) \approx e^{i \omega r_\ast}$ in the large-$r$ limit. The residue is approximately
\beq
\tilde{s}_{\omega n}^{\pm} \approx \pm 8 \left( - i \omega r_c \right)^N \left( \frac{64 r_c^2}{l_c^2} \right)^n \frac{e^{-2 i \Delta}}{(2 \pi)^{1/2} n!} ,  \label{res-approx}
\eeq
where
\beq
\Delta \equiv \omega r_c - \tilde{\omega} \frac{r_h}{2} \ln \left( \frac{r_c - r_h}{r_c + r_h} \right) .
\eeq

Orbiting oscillations in the cross section $|f_\omega|^2$ arise from interference between co-rotating $(+)$ and counter-rotating $(-)$ contributions to (\ref{cam-scat-ampl}). The angular oscillation frequency is given by $\text{Re} \left( m_{\omega n}^{+} - m_{\omega n}^{-} \right) \approx \omega ( l_c^{+} + \left| l_c^- \right| ) = 4 \omega r_e$ at lowest order (where $r_e = \sqrt{C^2+D^2} / c$). Therefore, the angular width of the orbiting oscillations at high-frequency is simply
\beq
\lambda_\phi \approx \frac{\pi}{2 \omega r_e} .
\eeq
The imaginary part of $m_{\omega n}^+$ (or $m_{\omega n}^-$) determines the rate of attenuation with scattering angle $\phi$ (or $2 \pi - \phi$). 

\subsubsection{Numerical results}
\begin{figure*}
\includegraphics[width=8.2cm]{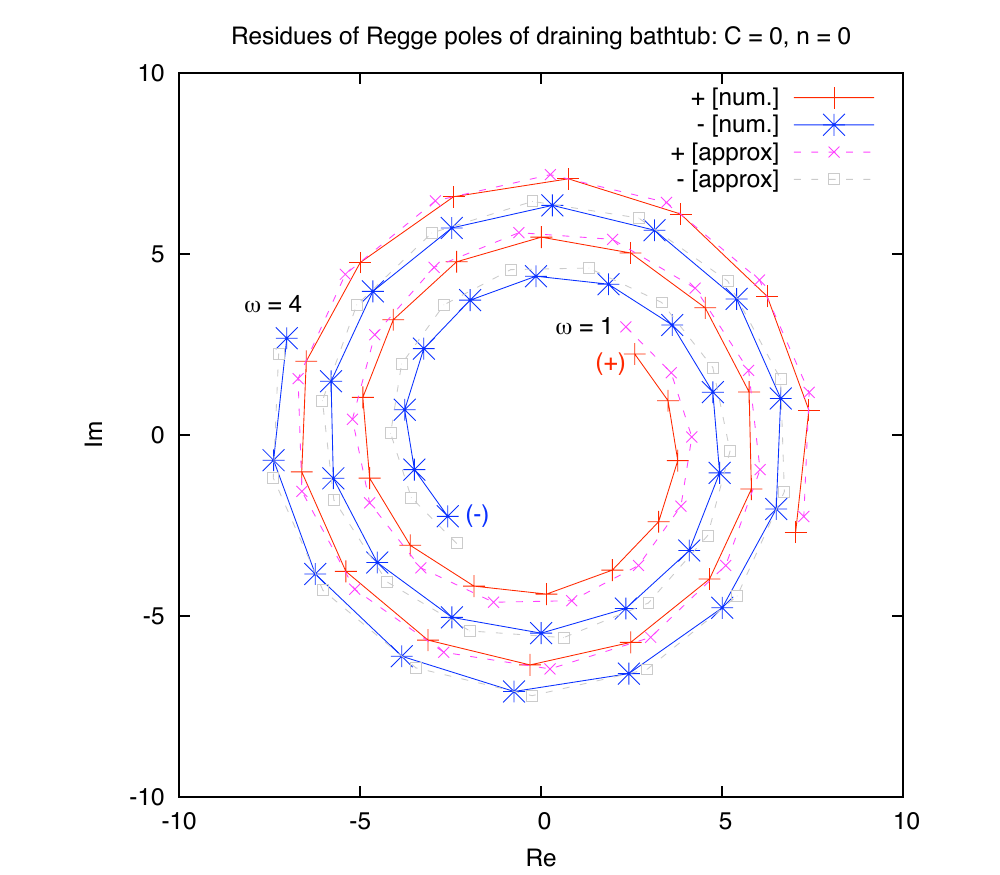}
\includegraphics[width=8.2cm]{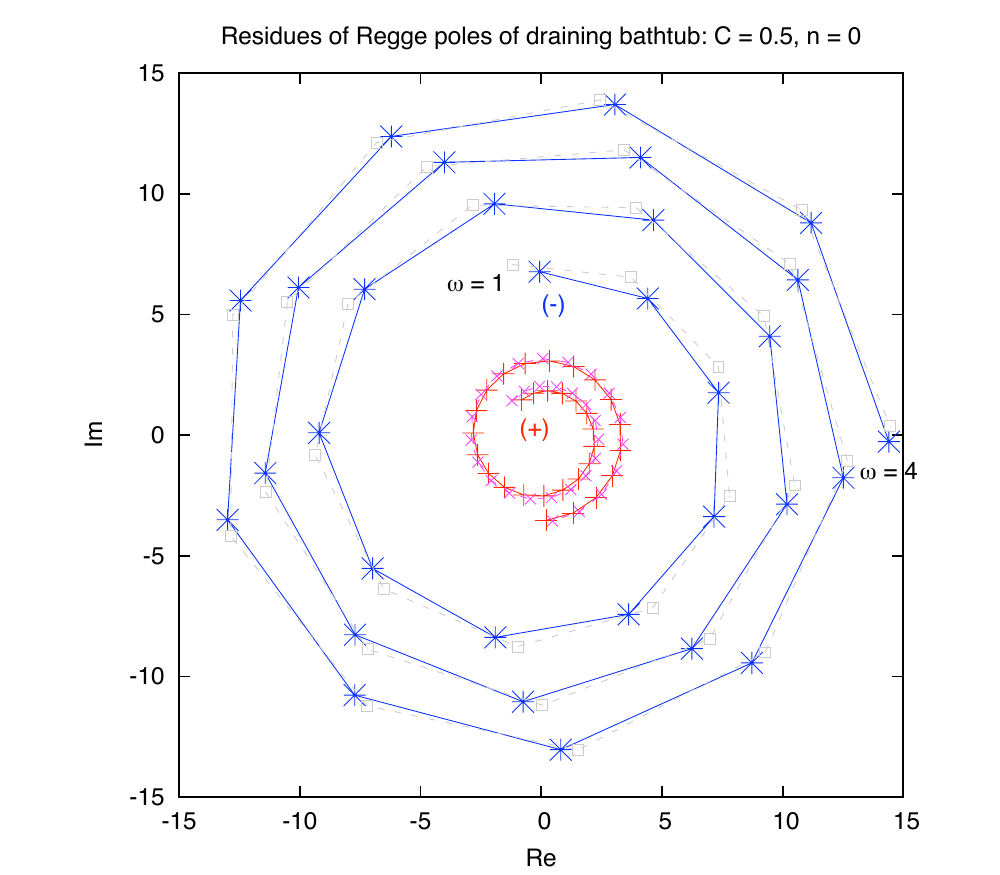}
\caption{Regge pole residues $\tilde{s}_{\omega m}^{\pm}$ in complex plane. The plots show the residues defined in Eq.~(\ref{rp-res-def}) for the fundamental ($n = 0$) modes of a DBT, for the cases $C = 0$ (non-rotating) [left plot] and $C / D = 0.5$ [right plot]. The blue and red crosses (solid lines) show the numerically-determined residues for `prograde' (+) and `retrograde' (-) modes, for frequencies $\omega r_h = 1.0, 1.1, 1.2, \ldots, 4.0$. The dotted lines show the high-frequency approximation, Eq.~(\ref{res-approx}).}
\label{fig-rp-res}
\end{figure*}

Figure \ref{fig-rp-res} shows a comparison between the high-frequency approximation for the residues (\ref{res-approx}), and accurate values computed using a numerical method based on direct integration. We find a good agreement, suggesting that the analytic approximation Eq.~(\ref{res-approx}) can be used to capture the key features of orbiting. 

\begin{figure}
\includegraphics[width=8cm]{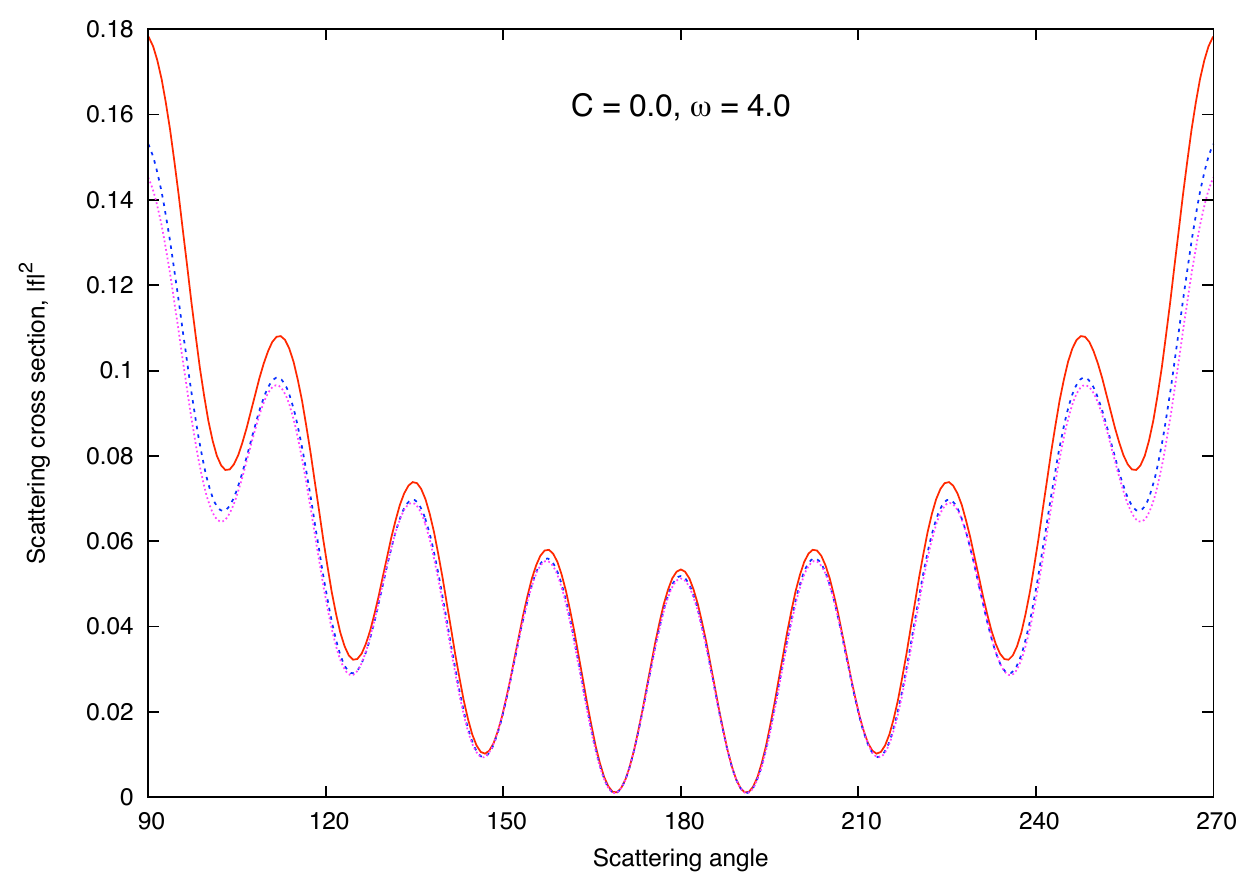}
\includegraphics[width=8cm]{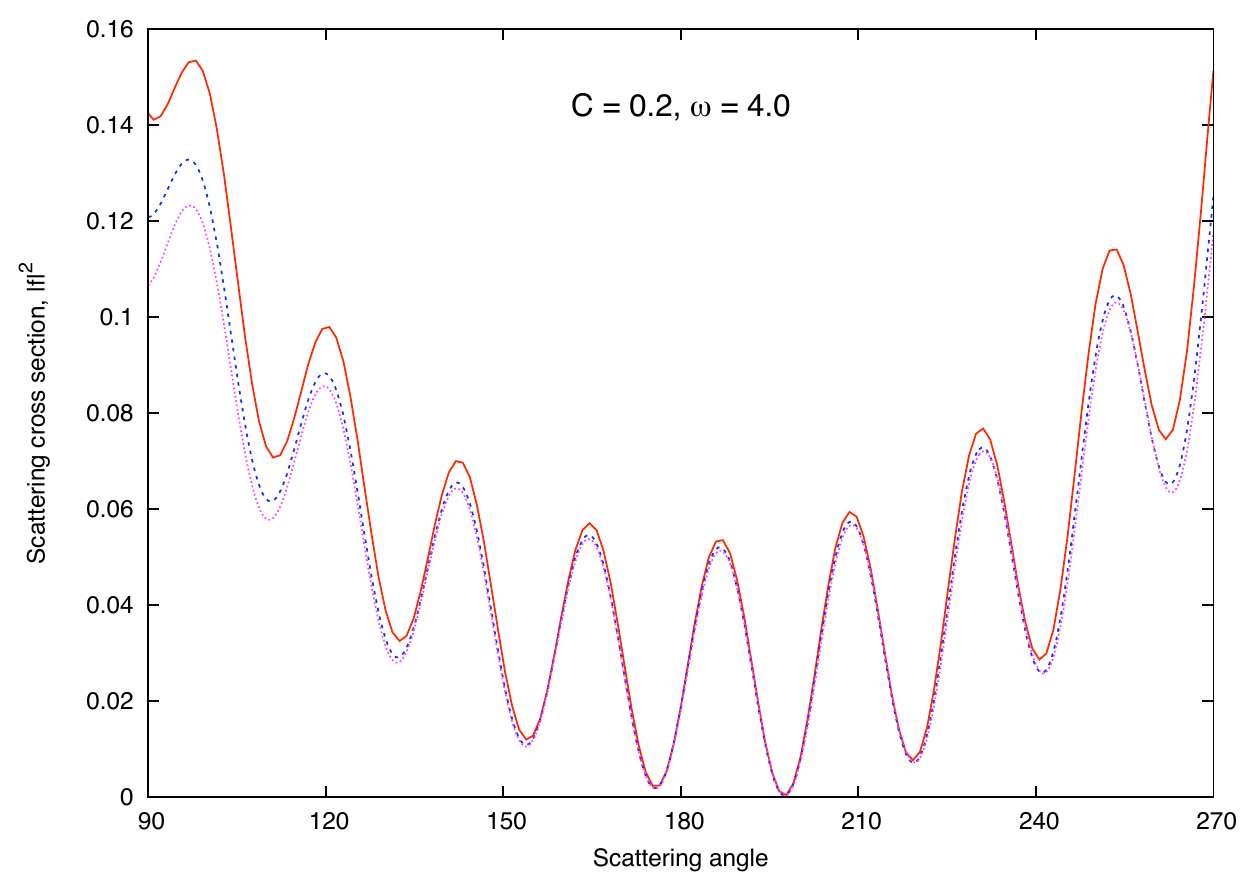}
\includegraphics[width=8cm]{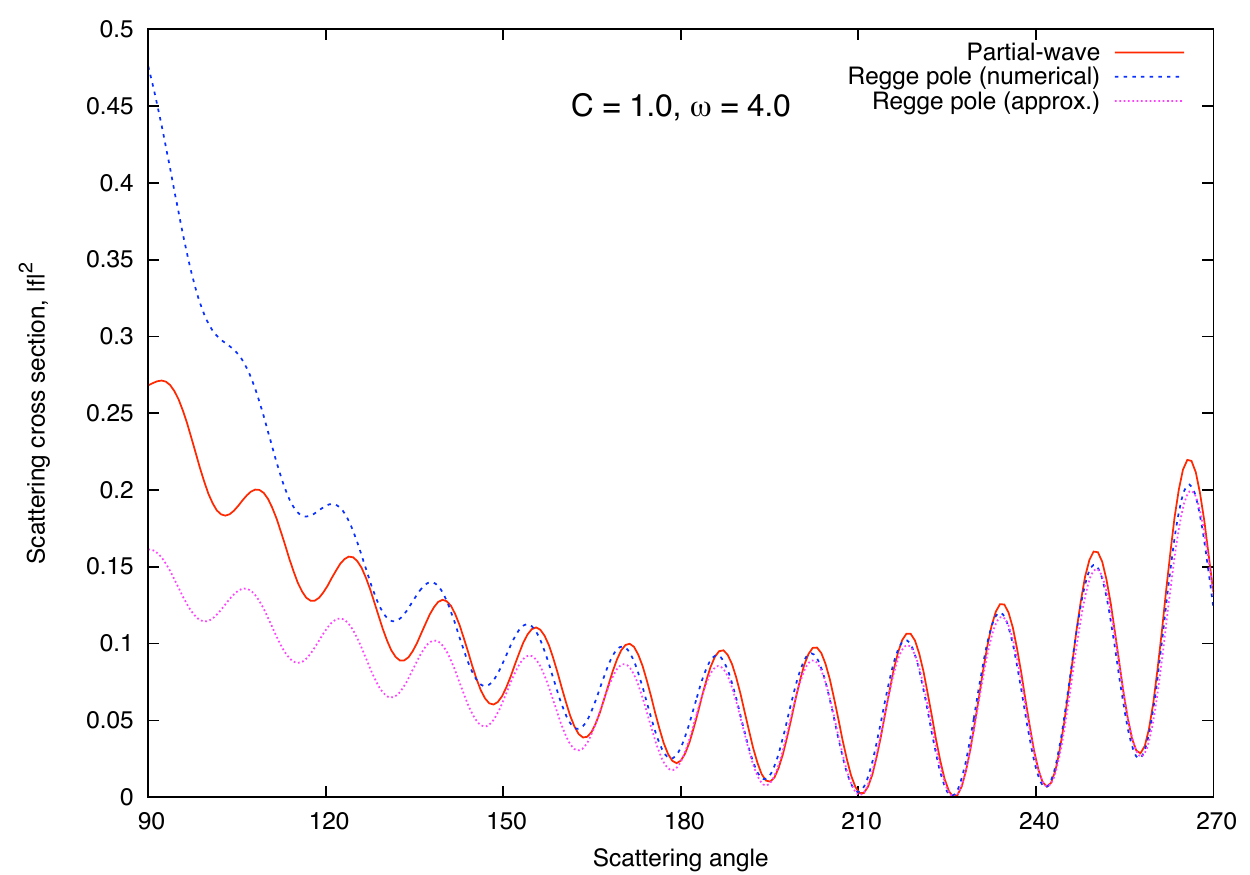}
\caption{Orbiting and the CAM approximation. These plots show the scattering cross section $d\sigma/d\phi = |f_\omega(\phi)|^2$ as a function of scattering angle, with a coupling $\omega r_h = 4$, for three cases: $C = 0$ (non-rotating), $C=0.2 r_h$ and $C=1.0 r_h$. The plots compare the cross section found from the partial wave series [red, solid] with the cross section found from the Regge pole (RP) approximation (\ref{cam-scat-ampl}) [dotted line], using only the lowest modes $n=0\ldots 4$). The blue dotted line shows the RP approximation with numerically-determined RPs and residues. The magenta dotted line shows the RP approximation using asymptotic expressions (\ref{momex}) and (\ref{res-approx}), for $n=0$ only. }
\label{fig-cam}
\end{figure}

Figure \ref{fig-cam} compares the CAM approximation for orbiting (\ref{cam-scat-ampl}), with accurate numerical results obtained by summing the partial wave series \cite{ODC2}. Two CAM approximations are shown; for the first approximation we computed the RPs and residues numerically (for low overtones $n=0\ldots4$), and for the second approximation we used the asymptotic analytic results, (\ref{momex}) and (\ref{res-approx}). 

The agreement is found to be good, particularly at large scattering angles. In other words, the Regge pole approximation, given in (\ref{cam-scat-ampl}), accounts rather well for the orbiting oscillations. 
In particular, the asymptotic results (given in closed form in terms of geodesic parameters $r_c^\pm$ and $l_c^\pm$) provide an excellent description in the semi-classical (high-frequency) regime. 
%


\section{Final Remarks\label{conclusions}}
In this paper we have focused our attention on the quasinormal frequencies and Regge poles of the draining bathtub, a simple system which may be studied in the lab and which has been suggested as a simple (though inexact) analogue for the Kerr spacetime \cite{Schutzhold-Unruh-2002}. Our aim was to bring together a variety of techniques, both old \cite{Cardoso-Lemos-Yoshida} and new \cite{Dolan-Ottewill, Dolan, DFR, Dolan-Ottewill-2011}, to build up a unified picture of the resonances of a simple non-spherically-symmetric system, for the first time. 

We studied the QN resonances numerically in both the time domain (Sec.~\ref{subsec:qn-timedomain}) and the frequency domain (Sec.~\ref{sec:qn-freq}). For the former we developed a purpose-built finite-difference code, and for the latter we applied the continued-fraction method (developed in Ref.~\cite{Cardoso-Lemos-Yoshida}). We provided a clear geometric interpretation (Sec.~\ref{subsec:geometric}) for our numerical results, in terms of the properties of the co- and counter-rotating photon/phonon orbits of the spacetime (Sec.~\ref{subsec:geodesics}). The geometric interpretation neatly explains the similarities and differences between the DBT and rotating black hole QN spectra (see Fig.~\ref{fig:qn-spectrum}). Furthermore, the geometric interpretation may be leveraged to obtain an expansion of QN frequencies in inverse powers of $m$, as we showed in Sec.~\ref{sec:expansion}.

In Sec.~\ref{RPs}, we demonstrated how methods developed to investigate QN modes may be simply adapted to investigate the Regge pole spectrum (see also e.g.~\cite{DFR}). The relevance of the Regge poles is revealed by the Complex Angular Momentum method~\cite{Nussenzveig}, which we explored in Sec.~\ref{sec:CAM}. Regge poles may be used to account for oscillations seen in the absorption and scattering cross sections.  
First, in Sec.~\ref{subsec:abs-csec}, we extended a method of \cite{DEFF, DFR} which reveals the role of Regge poles in fine-structure oscillations in the absorption cross section $\sigabs$. We believe this represents the first application of this work to a non-spherically symmetric spacetime. We showed (Fig.~\ref{fig-abs}) that the superposition of diffraction effects linked to the properties of co- and counter-rotating orbits creates a rather irregular fine-structure \cite{DFR} in $\sigabs(\omega)$ when the system is rotating.
Next, in Sec.~\ref{subsec:scat-csec}, we applied the CAM method to obtain a geometric approximation for the `orbiting' oscillations in the scattering cross section. To make this possible, we used the geodesic expansion method \cite{Dolan-Ottewill-2011} to develop new approximations for the \emph{residues} of the scattering matrix (Appendix \ref{Appendix:RP-residues}), which are valid in the high-frequency limit. 

Let us briefly highlight some important analytic results herein: (i) at late times, perturbations of the DBT undergo power-law decay with an index $\eta$ given by Eq.~(\ref{power-law-index}); (ii) the isotropic ($m=0$) mode possesses a simple analytic solution, Eq.~(\ref{uin-m0}) and (\ref{uup-m0}); (iii) asymptotic expansions for QN and RP frequencies in terms of geometric quantities are given by Eq.~(\ref{freqex}) and (\ref{momex}), and they are good approximations in the large-$|m|$ and large-$|\omega|$ regimes, respectively (Fig.~\ref{fig:validation}); (iv) the absorption cross section may be expressed in terms of Regge poles via Eq.~(\ref{sig-cam}); (v) at large frequencies, the simple `geometric' formula Eq.~(\ref{abs-geo}) provides a good approximation to $\sigabs$ (see Fig.~\ref{fig-abs}); (vi) the scattering amplitude at large angles is approximated by the Regge pole formula (\ref{cam-scat-ampl}); (vii) in the high-frequency limit, the RP residues [required for (vi)] are given by Eq.~(\ref{res-approx}).

In conclusion, we have illustrated that a simple analogue system (the DBT) possesses a spectrum of resonances which, in addition to being of interest in their own right, can improve our understanding of the resonances of a rotating black hole. Whereas black holes are remote celestial entities, analogues may be investigated in the lab today \cite{Wein}. We hope that a laboratory investigation of the resonances of the DBT will soon be undertaken. Here, experimentalists face the challenge of maintaining the stability of a \emph{converging} flow, particularly at the point at which the flow becomes supersonic. It may be easier to instead use a \emph{diverging} supersonic flow, which would instead represent a \emph{white hole} analogue. For example, an hydraulic jump~\cite{Jannes1} extended to a rotating set-up may be of interest. It has to be noted, though, that the boundary conditions used to determine resonances of a white hole are different from the ones used in the corresponding black hole case~\cite{Nigel, Jannes2}. A further avenue for investigation is the effect of high-frequency dispersion on the resonance spectrum. For example, it has been argued that superluminal dispersion creates continuous zones in the QN mode spectrum of Bose-Einstein condensates~\cite{Jannes3}. Dispersion would certainly affect the validity of the short-wavelength (i.e.~large-$m$ and large-$\omega$) approximations developed here, and would presumably disrupt the direct link between the properties of orbiting geodesics and the spectrum of resonances. Of further interest are `pinned vortices' in superfluids, which resemble (to a certain extent) three-dimensional but non-draining versions of the DBT; it has recently been suggested that here bound states propagating along the vortex (similar to  `whispering gallery' modes) may arise \cite{Marecki-Schutzhold}.

To summarise, the key message from the present work is that the presence of distinct prograde and retrograde null geodesic orbits on a rotating (effective) spacetime leaves its clear imprint on the resonance spectrum. In turn, the superposition of co- and counter-rotating resonances generates a host of potentially-observable effects, for example (i) in QN ringing; (ii) in the fine-structure of the absorption cross section; and (iii) in the  `orbiting' diffraction effect in scattering. Here, we have applied a range of methods to investigate (i)--(iii) for the rather `idealized' geometry of the draining bathtub. It is our hope that these methods can now be extended to investigate rotating analogue systems of current experimental interest.

\appendix
\section{Regge pole residues\label{Appendix:RP-residues}}
In this section we apply the geodesic expansion method to obtain an asymptotic approximation for the Regge pole residues which is valid in the high-frequency limit. The method follows the outline given in Appendix A of Ref.~\cite{Dolan-Ottewill-2011}, where a similar calculation was performed for the (spherically-symmetric) Schwarzschild black hole. Here, the situation is complicated somewhat by the rotation of the system ($C \neq 0$). 

Recall that the residue is defined by Eq.~(\ref{rp-res-def}). Here we will use the overbar notation $\bar{m} = m^{\pm}_{\omega n}$ to represent the Regge pole value.
To find the residue, we perturb the azimuthal number slightly away from its Regge pole
value, $m \equiv \bar{m} + \epsilon$, (where $\epsilon$ is small) to determine the
first-order change in $\Ain$. Away from the RP value, the `global' wavefunction ansatz (\ref{mode1}) is no longer valid: its continuity breaks down near $r \approx r_c$. Instead, we may find regular solutions in
a ÔÔinteriorÕÕ region close to $r = r_c$ and match these onto
ÔÔexteriorÕÕ solutions (in regimes $r \lesssim r_c$ and $r \gtrsim r_c$),
to obtain $\Ain = \mathcal{O}(\epsilon)$ to first order. In other words, we combine the geodesic expansion method with a standard WKB approach.

\subsection{Interior solution}
Let us begin with the radial equation (\ref{radeq}) and first make the substitution $\uin_m = f^{-1/2} \uint(r)$, to obtain
\beq
\frac{d^2 \uint}{dr^2} + \mathcal{U}(r) \uint = 0  ,  \label{uint-eq}
\eeq
where
\beq
 \mathcal{U}(r) =  f^{-2} \left[ \left(\omega - \frac{C m}{r^2}\right)^2 - f \left( \frac{m^2 - 1/4}{r^2} - \frac{7D^2}{4r^4} \right) + \frac{D^4}{r^6} \right] . \nn
\eeq
Now, to find a solution valid in the vicinity of $r=r_c$, we make the change of variables $r = r_c + \rho \, \omega^{-1/2} z$, where
\beq
\rho \equiv \frac{\lc}{4} \sqrt{\frac{2}{\rc}} .  \label{rhodef}
\eeq
Substituting (\ref{rhodef}) into (\ref{uint-eq}), with the assumption that $\omega$ is large, brings us to the following equation,
\beq
\frac{d^2 \uint}{dz^2} + \left[ z^2 - 2 i \left( n + \eta + 1/2 \right) + \mathcal{O}\left(\omega^{-1/2}\right) \right] \uint = 0 ,  \label{eq-interior}
\eeq
where $\eta \equiv - i \epsilon r_c / l_c$. Note that in our derivations we make use of the following identities: $\rc^2 - C \lc = \tfrac{1}{2} \lc^2 = 2 (r_c^2 - D^2) = 2 (D^2 - C \lc)$. The solutions of Eq.~(\ref{eq-interior}) are the parabolic cylinder functions $D_{n+\eta}((-1+ i) z)$ and $D_{n + \eta} ((1 - i)z)$. We use only the former solution, because (as we shall see) it matches onto an `exterior' solution which is purely ingoing at the horizon. That is, we choose
\beq
\uint = D_{n + \eta} \left( (-1 + i) z \right),
\eeq

\subsection{Exterior solutions}
In the exterior regimes $r \lesssim r_c$ and $r \gtrsim r_c$, we require a pair of solutions $u^{\pm}(r)$ which are (i) approximate solutions to the radial equation in the large-$\omega$ limit, and (ii) have well-defined `ingoing'/`outgoing' behaviour in the limits $r_\ast \rightarrow \pm \infty$. To obtain these solutions, we start with an ansatz of the form (\ref{mode1}), specifically,
\beq
u_m^{\pm} = \chi^{\pm}(r) \exp \left[ \pm i \int_{r_c}^r \alpha(r') f^{-1}(r') dr' \right] 
\eeq
with $\alpha$ as defined in Eq.~(\ref{alp-def}), and $\chi^{\pm}(r)$ of the form
\beq
\chi^{\pm} \approx \left(1 - \frac{r}{r_c} \right)^n \exp(S_0^\pm(r)) .
\eeq 
To find explicit solutions, we expand the exponent $\alpha$ at orders (i) $\omega^1$ and (ii) $\omega^0$, and (iii) obtain $S_0^\pm$ by solving an equation like (\ref{order}). After these steps (and finding some cancellation between terms at stages (ii) and (iii)), we reach the result
\beq
u^{\pm}_m(r) = \left( \frac{|r - r_c|}{r + r_c} \right)^{\pm N - 1/2} \left( \frac{r}{r+r_c} \right) e^{\pm i \omega r} \left(\frac{r - D}{r + D} \right)^{\mp i \tilde{\omega} D / 2} ,
\eeq
where $\tilde{\omega} = \omega - m C / D^2$.

\subsection{Matching}
Let us now introduce three solutions,
\beq
\begin{array}{l l l}
u_{>} &=  a^+ u^+  + a^- u^-, \quad & r \gtrsim r_c \\
u_0 &= D_{n + \eta} ((-1+i) z) , \quad & r \sim r_c \\
u_{<} &= b^{+} u^{+} , \quad & r \lesssim r_c .
\end{array}
\eeq
where $a^+$, $a^-$ and $b^+$ are coefficients to be determined. 
To match these solutions together in the transition regions, we make use of the following asymptotic forms:
\begin{widetext}
\beq
u_0 \sim \begin{cases} 
2^{n/2} e^{-i \pi n /4} |z|^n e^{+i z^2 / 2},  & z \rightarrow -\infty , \\  
2^{n/2} e^{3 i \pi n / 4} |z|^n e^{+i z^2 / 2}  
+ \eta \Gamma(n+1) (2\pi)^{1/2} |z|^{-(n+1)} 2^{-(n+1)/2} e^{-3 i \pi (n+1) / 4} e^{-iz^2/2},
& z \rightarrow + \infty  .
 \end{cases}
\eeq
\end{widetext}
and
\beq
u^{\pm} \sim  A F^{\pm} \left(B |z| \right)^{\pm N - 1/2} e^{\pm i z^2 / 2} ,
\eeq
with $A = 1/2$, $B = \rho / [2 r_c \omega^{1/2} ]$, and
\beq
F^{\pm} = \exp \left( \pm i \left[ \omega r_c - \frac{\tilde{\omega} D}{2} \ln \left( \frac{r_c - D}{r_c + D} \right) \right] \right).
\eeq
By matching the asymptotic forms, it follows immediately that
\beq
\frac{\Aout}{\Ain} = \frac{a^+}{a^-} = \frac{(F^-)^2 B^{-2N} 2^{N} e^{3i\pi N / 2}}{\eta (2\pi)^{1/2} \Gamma(n + 1)} .
\eeq
This leads us to the final (approximate) result for the residue, given in Eq.~(\ref{res-approx}). 

\begin{acknowledgments}
The authors would like to thank Conselho Nacional de Desenvolvimento 
Cient\'\i fico e Tecnol\'ogico (CNPq) and Funda\c{c}\~ao de 
Amparo \`a Pesquisa do Estado do Par\'a (FAPESPA)
for partial financial support. 
S.~D.~thanks the Universidade Federal do Par\'a (UFPA) 
in Bel\'em for kind hospitality, and acknowledges financial support from the Engineering and Physical
Sciences Research Council (EPSRC) under grant no. EP/G049092/1.
L.~C.~and L.~O.~would like to acknowledge also partial financial 
support from Coordena\c{c}\~ao de Aperfei\c{c}oamento de Pessoal
de N\'\i vel Superior (CAPES). L.~C.~thanks the Abdus Salam International Centre for Theoretical Physics (ICTP) for kind hospitality. The authors thank Ednilton Oliveira for helpful discussions.
\end{acknowledgments}


\begin{thebibliography}{99}
\bibitem{footnote}
Elsewhere in the literature, $C$ and $D$ are alternately referred to as $B$ and $A$, respectively.
\end{thebibliography}
\end{document}